\documentclass[twoside,twocolumn,prd,amsmath,superscriptaddress,eqsecnum,nofootinbib,showpacs]{revtex4}

\usepackage{amssymb}
\usepackage{graphicx}

\usepackage{fancyhdr}
\renewcommand{\sectionmark}[1]{\markboth{\textsc{Bi\v c\'ak} {\it et al.}}{\textsc{Effects of rotating gravitational waves}}{}}

\begin{document}
\fancyhead[RO]{\thepage}
\fancyhead[LE]{\thepage}

\fancyhead[RE]{{\footnotesize\rightmark}}
\fancyhead[LO]{\footnotesize{\leftmark}}

\title{Effects of rotating gravitational waves}
\author{Ji\v r\'\i ~Bi\v c\'ak}
\email{bicak@mbox.troja.mff.cuni.cz}
\affiliation{Institute of Theoretical Physics,  Charles University, 180 00 Prague 8, Czech Republic}
\affiliation{Institute of Astronomy, Madingley Road, Cambridge CB3 0HA,United Kingdom}
\affiliation{Albert Einstein Institute, Am M\"uhlenberg 1, D-14476 Golm, Germany}

\author{Joseph Katz}
\email{jkatz@phys.huji.ac.il}
\affiliation{The Racah Institute of Physics, Givat Ram, 91904 Jerusalem, Israel}
\affiliation{Institute of Astronomy, Madingley Road, Cambridge CB3 0HA,United Kingdom}

\author{Tom\'a\v s Ledvinka}
\email{ledvinka@mbox.troja.mff.cuni.cz}
\affiliation{Institute of Theoretical Physics,  Charles University, 180 00 Prague 8, Czech Republic}

\author{Donald Lynden-Bell}
\email{dlb@ast.cam.ac.uk}
\affiliation{Institute of Astronomy, Madingley Road, Cambridge CB3 0HA,United Kingdom}


\begin{abstract}
We study gravitational waves to first and second order in amplitude in vacuum asymptotically flat spacetimes. The Einstein equations  are solved to first order and these solutions are superposed to form a time-symmetric  ingoing and then outgoing pulse regular everywhere. The waves are assumed to have odd-parity and a non-vanishing angular~momentum  which keeps them away from the axis at all times. The averaged energy of the waves is evaluated. The relevant Einstein equation  is then solved to second order in the amplitude. The influence of the angular~momentum  of the waves on the rotation of local inertial frames  with respect to the frames at great distances is analyzed. The rotation of the frames  occurs even in the region around the origin where spacetime  is almost flat. The rotation is without time delay as it follows from the constraint equation.

The results are illustrated graphically for various values of the ``harmonic  index $m$" corresponding to azimuthal rotation and the ``harmonic index $l$'' describing the latitudinal rotation of the waves. The apparent motions of the fixed stars on the celestial sphere as seen through rotating waves from the local inertial frame are calculated and displayed.
\end{abstract}
\maketitle

\section{Introduction}
It was just 100 years ago in Prague when Einstein wrote the paper \cite{Ei} in which he, for the first time, expressed his understanding of Mach's Principle. Within his pre-general relativity  theory in which there was only one metric function he considered a mass point inside a shell accelerated ``upwards" and found that the mass point is dragged along by the shell.

Many formulations and studies of Mach's Principle have appeared during the last 100 years, most of them were analyzed in the T\"ubingen conference in 1993 which led to the remarkable volume \cite{BP} containing lectures as well as valuable discussions. We studied Machian effects in various contexts, both in asymptotically flat spacetimes 
 and within cosmological perturbation  theory -- see, e.g., \cite{BKL07}, and number of references therein; later, cf. Schmid \cite{CSchmid}.

More recently, we investigated the subtle question whether dragging of inertial frames  should also be attributed also to gravitational waves. After the discovery of binary pulsars losing energy and angular momentum as a consequence of emitting
 gravitational radiation it would be surprising if gravitational waves did not have an influence on local inertial frames. However, there are still doubts about the status of gravitational stress-energy as compared with stress-energy tensor $T_{\mu\nu}$ of matter in relation to Machian ideas (see, e.g., \cite{BP}, p. 83).

 In \cite{BKL08cqg1} and \cite{BKL08cqg2}, we studied gravitational waves with translational symmetry that rotate about the axis of cylindrical polar coordinates. We solved the Einstein equations to first order in the wave amplitude and formed a pulse. By solving the relevant Einstein equation  to second order we calculated and illustrated graphically the rotation of a local  inertial frame  at the axis which is caused by the rotating wave pulse.

 In the present work we investigate the effects of rotating gravitational waves  in a more general, asymptotically  flat setting, without assuming any symmetry. We again start out from linearized theory and construct an ingoing rotating pulse of radiation which later transforms into an outgoing pulse. While in the translational symmetric case our waves were characterized by just one harmonic index $m$ governing the number of wave crests in $\varphi$, now the situation becomes considerably richer involving both spherical harmonic indices $l$ and $m$.

 Our problem now resembles, in some respects, the study by Lindblom and Brill \cite{LiBr} of the gravitational collapse of a slowly rotating, massive spherical shell freely falling under its own gravity. We reconsidered and extended their analysis in \cite{KLB98}. The construction of our waves starts out from (the linearized version of) a Bonnor-Weber-Wheeler time-symmetric, ingoing and outgoing pulse of a typical width $a$ and a typical amplitude $C$. One can make a pulse that is rotating, but still localized in the radial direction so that it resembles a  falling and rotating shell. Although we do not find the complete metric up to the second order in the amplitude, the resulting spacetime  will be asymptotically  flat at spatial infinity since the first-order perturbations  decay as $h_{\mu\nu}^{(1)}\thicksim r^{-(l+1)}$ there, $l$ being the multipolar index. Hence, one can define ``fixed stars" at infinity at rest in an asymptotically  Minkowski system given in asymptotically  spherical coordinates $t, r,
 \theta, \varphi $. The waves have mass-energy, a nonvanishing total angular momentum, and a vanishing total linear momentum. Since we consider weak waves it is plausible to assume that a global coordinate system can be extended over the whole spacetime. At infinity it becomes the rest frame of the fixed stars. 

Near the origin, the first-order metric of our waves behaves as $r^l$, so the region around the origin will be very nearly flat for $l$ sufficiently large. When, however, a local inertial frame  is introduced at the origin, we find that its axes rotate with respect to  the lines $\varphi=\rm const$ of the global frame, i.e., with respect to fixed stars at infinity.

 Near the origin the congruence of the worldlines $\varphi=\rm const$ twists and the observers attached to these lines experience Euler acceleration proportional to $d_t \omega_0$, where $\omega_0$ is the angular velocity of the inertial frame  near the origin. 
 The angular velocity  $\omega_0$ enters {\it the second-order odd-parity dipole $l=1$ perturbation } of the metric, $g_{t\varphi }^{(2)}= - \omega_0\; r^2\sin^2\theta$. (The centrifugal acceleration is higher order in the angular velocity.) The situation thus indeed resembles the interior of a collapsing slowly rotating shell -- see \cite{KLB98} where the vorticity of the lines $\varphi=\rm const$ is given in covariant form. In \cite{KLB98} we also calculated how the fixed  stars at infinity rotate with respect to  the inertial frame  at the origin by considering photons emitted radially inwards from the stars.

 In the present work we first consider gravitational waves  with odd parity in the Regge-Wheeler gauge in the linearized Einstein theory. The metric components $h_{\mu\nu}^{(1)}$ describing the waves can all be derived from one potential-type function satisfying the flat-space scalar wave equation. We choose this function in the form of a generalized Bonnor-Weber-Wheeler ingoing and outgoing pulse characterized by specific harmonic multipolar orders $l$ and $m, |m|\le l$. For $m\ne0$, the field represents the superposition of {\it rotating} waves which are ingoing from infinity towards the origin, stay regular all the time, and become outgoing waves again to decay at infinity.

 Turning to the second-order metric $h_{\mu\nu}^{(2)}$, we employ the second-order perturbative Einstein equations  which are well known to be again linear differential equations  that contain as ``source terms" expressions quadratic in the first-order perturbations  $h_{\mu\nu}^{(1)}$ and their first- and second-order derivatives (see, e.g., \cite{NiGP}, \cite{GNPP}). Since we look for the second-order rotational perturbations  we concentrate on odd-parity again. To determine the influence of gravitational waves  on the rotation of local inertial frames  at the axis of symmetry we do not need to solve the second-order Einstein equations  in general, because the high $l$ terms do not have significant amplitude at the origin. The terms that give the inertial frame near the origin are $l=1, m=0$, so we can average the ``source terms" over azimuth $\varphi $, and the problem becomes axisymmetric. This simplifies the calculations. In addition, an important result proved and discussed in \cite{NiGP}, \cite{GNPP} 
states that the Regge-Wheeler gauge can be introduced in both first and second order, and it is unique. As emphasized in \cite{GNPP}, perturbations  in the Regge-Wheeler gauge can be considered to be gauge-invariant expressions.

 The angular velocity  $\omega_0$ of rotation of the local inertial frames at the origin, whose neighborhood remains as flat as we wish for all times by choosing very high $l$, is found by solving a simple linear elliptic equation  for the second-order dipole odd-parity perturbation  with a complicated source term given by the averaged component, $\langle R_{t \varphi}^{(2)}[h^{(1)}, h^{(1)}]\rangle $, of the Ricci tensor, quadratic in $h^{(1)}$ and its derivatives.

 A lengthy, complicated form of $\langle R_{t \varphi}^{(2)}\rangle $ allows us to find $\omega_0$ explicitly in a useful form only at particular values of time. However, by using computer algebra system we produce nice ``bell-shaped" curves showing how $\omega_0$ increases as the waves approach the origin and decreases as the waves recede back to infinity. From $\omega_0$  plotted for different azimuthal harmonic index $m$, with $l$ fixed, it is easily seen how an increase of the azimuthal angular~momentum  of the waves implies a stronger dragging of inertial frames  (gyroscopes) in nearly flat regions surrounded by the waves.

 We also calculate how the celestial sphere
will look when observed from the inertial frame  at the origin. This is influenced both by the passage of the stellar light through gravitational waves  and by the dragging of the inertial frame  due to their rotation. The resulting figures describing the apparent motion of fixed stars on the celestial sphere nicely illustrate the role of the multipolar harmonic indices $l$ and $m$ on the pattern of the motion.

The paper is organized as follows.
Since odd-parity perturbations  follow from a function satisfying the wave equation, in the next Sections II and III we study rotating scalar-field waves in Minkowski space. We evaluate their energy density and angular~momentum  density averaged over the azimuth $\varphi $ and in every direction, after averaging also over the latitude $\theta$.
We turn to rotating linearized gravitational waves  in Sections IV and V. Recalling some basic properties of the Regge-Wheeler formalism and tensor harmonics, and fixing our convention, we calculate the averaged energy density of the rotating linearized gravitational waves  and compare it with that of scalar waves. We find that for ``rapidly rotating" waves with $m\thicksim l$ the scalar field part dominates.

The core of the paper lies in Sections VI and VII. Here we consider the second-order perturbations. After expanding in tensor spherical harmonics, and taking the axially symmetric part, we find the second-order Einstein equations  for the radial parts of the second-order odd-parity metric components in the Regge-Wheeler gauge. For general $l$ we get linear wave-type equations  with source terms given by combinations of averaged components $\langle R_{\mu\nu}^{(2)}[h^{((1)}, h^{(1)}]\rangle $. For the dipole perturbations  we obtain an elliptic equation which we solve by variation of constants. The angular velocity   $\omega_0$ of an inertial frame  near the origin is determined and analyzed both analytically and numerically in Section VII. Here also the figures demonstrating the dragging of inertial frames
by the waves and the apparent motion of fixed stars as seen from the inertial frame at the origin are presented.  
In the concluding remarks we discuss the global, instantaneous character of the dragging of inertial frames as it follows from our results and show how it can be seen in diverse situations, including cosmological perturbations. We also indicate some open questions.
Some lengthy expressions are written down in Appendices. There, also various integrals over products of Legendre functions and non-trivial radial integrals are evaluated.


\section{Rotating scalar waves}
 In linearized Einstein theory the basic Regge-Wheeler equation \cite{RW} for odd parity waves becomes the usual flat-space 1-dimensional wave equation  for the radial part. It is thus natural to first present rotating waves in Minkowski  space.

 Consider the wave equation 
 \begin{equation}
 - \frac{\partial^2\psi}{\partial t^2}+ \Delta \psi=0
 \label{s2eq1}
 \end{equation}
in spherical coordinates. Assume
\begin{align}
\psi_{lm}&=
\frac{1}{2} \left[Q_l(t,r) Y_{lm}(\theta,\varphi )+\overline Q_l(t,r) \overline Y_{lm}(\theta,\varphi ) \right],
\label{s2eq2}
\\
Y_{lm}&=
N^m_l P^m_l(\cos\theta)e^{im\varphi },~~~l\ge0,~~|m|\le l;
\label{s2eq3}
\end{align}
bar denotes complex conjugation, $Y_{lm}$ are scalar spherical harmonics satisfying the standard normalization and orthogonality conditions over a unit sphere (see, e.g., \cite{Ja}),
\begin{align}
N^m_l = \sqrt{\frac{2l+1}{4\pi} \frac{(l-m)!}{(l+m)!}}~.
\end{align}
The angular part of Eq.$\,$(\ref{s2eq1}) then separates and the radial part satisfies
\begin{equation}
-\ddot Q_l+Q''_l+\frac{2}{r}Q'_l - \frac{l(l+1)}{r^2}Q_l=0.
\label{s2eq4}
\end{equation}
Hereafter the dot denotes $\partial_t$, the prime $\partial_r$.

\begin{figure}[hb]
\begin{center}
\includegraphics[angle = 0, width=6.4cm,bb=159 133 1052 1193,clip=true]{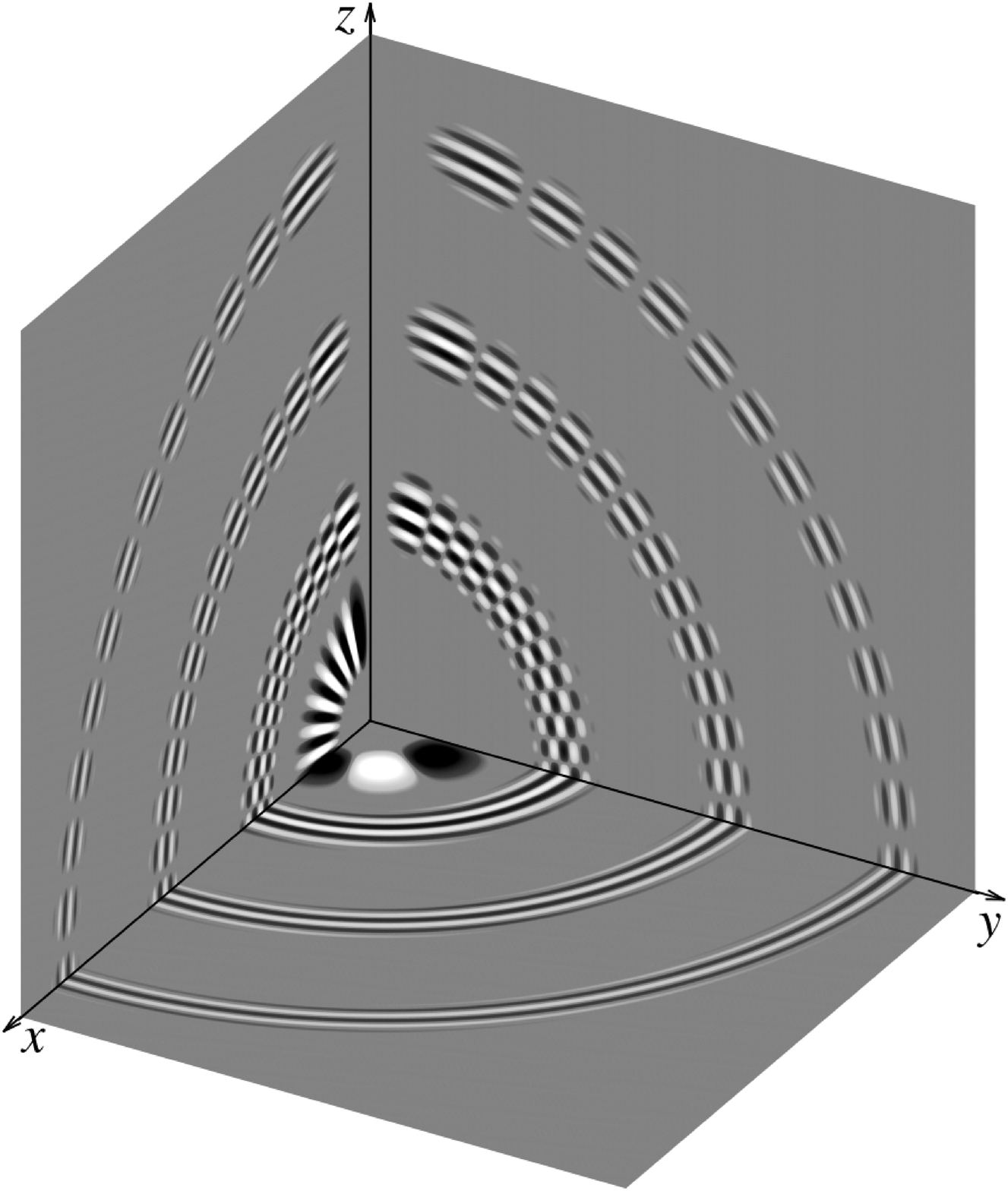}
\\
\includegraphics[angle = 0, width=6.4cm,bb=159 133 1052 1193,clip=true]{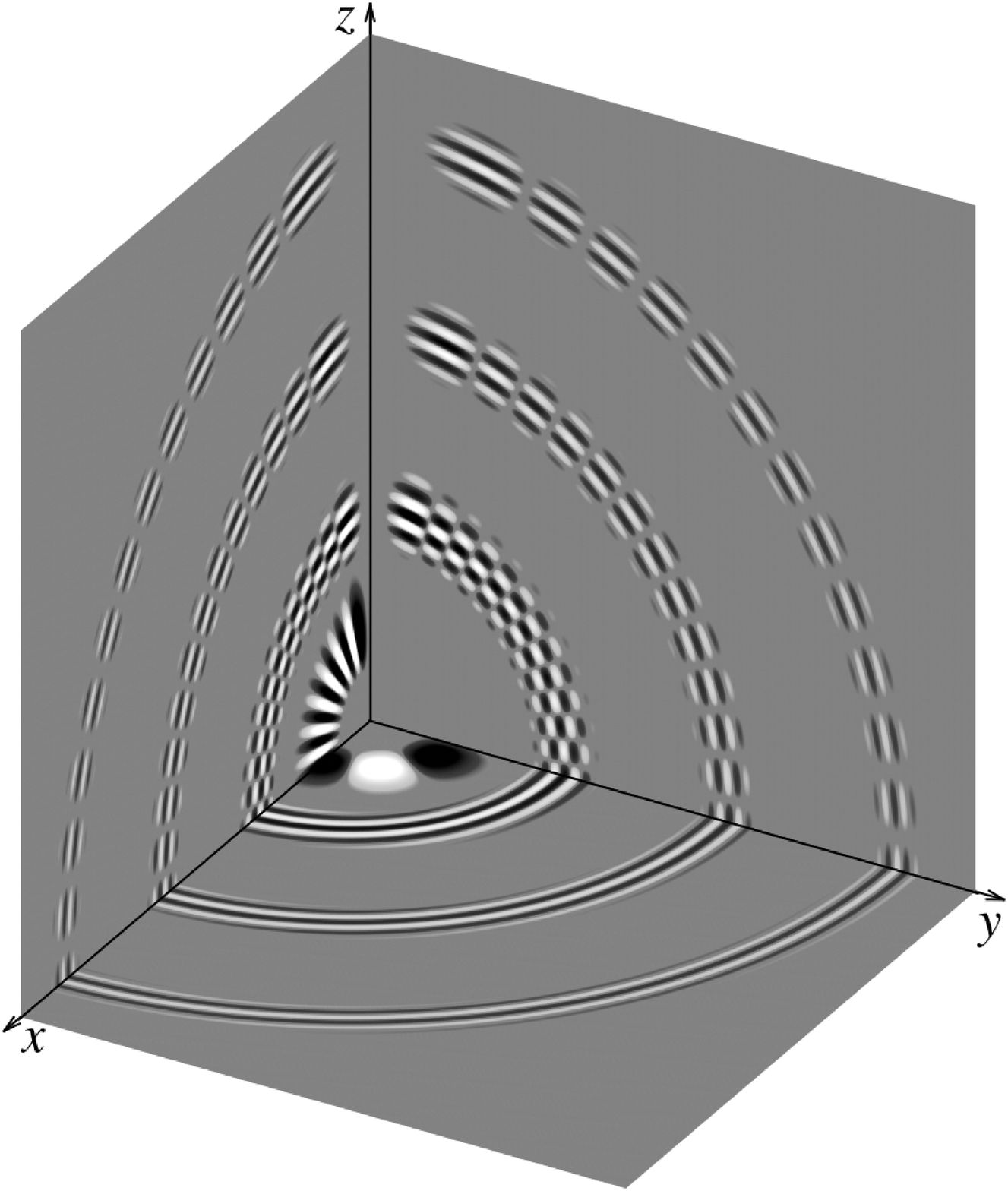}
\end{center}
\caption{\label{Figure1} Snapshots of function $\psi_{lm}$ for $l=27, m=5$, and times $\tilde t=-6,-4,-2, 0$ (upper image) or $\tilde t=0, 2, 4, 6$ (lower image).
For high values of $l$ the wave is concentrated near a shell with radius $r^2=a^2+t^2$, 
so we can combine waves at several moments into one figure.
The wave rotates anticlockwise around the $z$ axis. A careful observation reveals that the wave comes inwards in the form of a leading spiral, i.e.,
with the outside ahead of the inside (see ``white arms'' at $\tilde t=-2$ and $-4$); at $\tilde t=0$ the spiral structure 
has changed to a cartwheel but rotation keeps the wave away from the origin; at $\tilde t>0$ the spiral becomes trailing
(see "white arms" at $\tilde t=2$ and $4$). At large $\tilde t$ the spiral becomes tighter (cf. snapshots at $\tilde t=6$).
}
\end{figure}

Assume that $Q_l(t,r)=\hat Q_l(\omega,r)e^{i\omega t}$ and substitute
\begin{equation}
\hat Q_l(\omega,r)=\frac{1}{\sqrt{r}}u_l(\omega,r).
\label{25}
\end{equation}
Equation (\ref{s2eq4}) is then transformed into
\begin{equation}
u''_l+\frac{1}{r}u'_l - \frac{(l+{\textstyle{\frac{1}{2} }\!})^2}{r^2}u_l+\omega^2u_l=0,
\label{26}
\end{equation}
which is the Bessel equation  with index $\nu=l+{\textstyle{\frac{1}{2} }\!}$. The independent solutions of (\ref{26}) are given in terms of spherical Bessel and Hankel functions $j_l(\omega r), n_l(\omega r)$ and $h_l(\omega r)$. We want solutions regular at the origin, so we construct our radial waves as superpositions of
\begin{equation}
j_l(\omega r)=\sqrt{\frac{\pi}{2\omega r}} J_{l+{\textstyle{\frac{1}{2} }\!}}(\omega r),
\label{27}
\end{equation}
$J_l$ is a Bessel function of the first kind. Inspired by the example of the Bonnor \cite{Bo}, Weber-Wheeler \cite{WW} cylindrical pulse and by our recent work \cite{BKL08cqg1}, \cite{BKL08cqg2}, in which time symmetric incoming and outgoing rotating waves are smooth and finite everywhere at all time, we consider, in spherical polar coordinates, the superposition
\begin{align}
\label{s2eq8}
Q_l(t,r)&=B_l  \sqrt{\frac{\pi a^3}{2r}}
\int_0^\infty(a\omega)^{l+{\textstyle{\frac{1}{2} }\!}}\;e^{-a\omega} \; e^{-i\omega t} J_{l+{\textstyle{\frac{1}{2} }\!}}(\omega r) \;d \omega,
\end{align}
where the real amplitude $B_l$ and the characteristic width of the pulse $a$ are constant.
It is convenient to introduce 
\begin{equation}
\alpha=\alpha(t)=a+it.
\label{29}
\end{equation}

Employing then formula (8.6.5) in \cite{BE} (see also formula 6.623 in \cite{GR}), we find the integral in (\ref{s2eq8}) to yield
\begin{equation}
Q_l(t,r)= B_l2^ll! \; \frac{(r/a)^l}{[(\alpha^2+r^2)/a^2]^{l+1}}.
\label{s2eq10}
\end{equation}

To construct {\it rotating} waves coming in and then going out in a time-symmetric manner we just multiply by $Y_{lm}(\theta,\varphi )$ containing the factor $e^{im\varphi}$. Clearly, the field is the superposition of rotating waves: it is apparent that for each $\omega$ the wave contains the factor
\begin{equation}
e^{im\varphi }e^{-i\omega t}=e^{im(\varphi -\frac{\omega}{m}t)}
\label{s2eq12}
\end{equation}
so that the wave pattern rotates with the rate $\Omega_p=\omega/m$. The rotation and the pulse character of the wave as well as its regularity is easily seen when $\psi_{lm}$ is written explicitly in real terms:
\begin{equation}
\psi_{lm} = \frac{B_l\, 2^l\,l!\, N^m_l P^m_l(\cos\theta)~\tilde r^l}{[((1+\tilde r^2-\tilde t^2)^2+4\tilde t^2]^{(l+1)/2}}\cos\left[  m\varphi  - \lambda(t,r)  \right],
\label{s2eq13}
\end{equation}
where
\begin{equation}
\tilde r=\frac{r}{a}
~,~~~
\tilde t=\frac{t}{a}
\label{s2eq14}
\end{equation}
and
\begin{align}
\label{s2eq15}
\lambda(t,r) &= (l+1) \arctan\frac{2\tilde t}{1+{\tilde r}^2-{\tilde t}^2} 
\\\nonumber
	&=  {\rm arg}~\left[{\tilde r}^2+(1+i{\tilde t})^2\right]^{(l+1)}.
\end{align}

Some snapshots of the rotating waves profiles are illustrated in Figure \ref{Figure1}.


\section{Energy and angular~momentum  in rotating scalar waves}
We start with some simple properties of scalar fields. The energy density of the scalar field can be derived from the scalar field Lagrangian density
\begin{equation}
 {\cal L}= - {\textstyle{\frac{1}{2} }\!}\sqrt{-\overline g}\,\overline g^{\mu\nu}\partial_\mu\psi\partial_\nu\psi,
\label{31}
\end{equation}
where we use the background  Minkowski  metric in spherical coordinates
\begin{equation}
\overline g_{\mu\nu}=\{1, -\overline\gamma_{kl}\}=\{1, -1, -r^2,-r^2\sin^2\theta\}.
\label{32}
\end{equation}
(Hereafter, the Greek indices $\lambda, \mu, \nu = 0, 1, 2, 3$, the Latin indices $i,j ,k = 1, 2, 3$; $l,m$ are reserved for harmonic indices associated with $Y_{lm}$.)
The energy density, the time-time component of the stress-energy-momentum  tensor
\begin{equation}
\sqrt{-\overline g}~T^{\mu\nu}=\frac{\delta{\cal L}}{\delta\overline g_{\mu\nu}},
\label{33}
\end{equation}
reads
\begin{equation}
\varepsilon_S={{\frac{1}{2} }\!}\!\left(\dot\psi^2+\overline\gamma^{jk}\overline\nabla_j\psi\overline\nabla_k\psi\right)\!,
\end{equation}
$\overline\nabla_k\psi $ are covariant derivatives associated with $\overline\gamma_{jk}$; for scalar $\psi$, $\overline\nabla_k\psi =\partial_k\psi$, so in spherical coordinates
\begin{equation}
\varepsilon_S={{\frac{1}{2} }\!}\!\left[  \dot\psi^2+\psi'^2+\frac{1}{r^2}(\partial_\theta\psi)^2 +  \frac{1}{r^2\sin^2\theta}(\partial_\varphi \psi)^2  \right]\!.
\label{s3eq5}
\end{equation}
The energy density satisfies the conservation law which can be written as
\begin{align}
\label{35}
&\frac{d}{dt}\int_V{\textstyle{\frac{1}{2} }\!}\!\left(\dot\psi^2+\overline\gamma^{jk}\overline\nabla_j\psi\overline\nabla_k\psi\right) dV
\\\nonumber
&=\int_V\overline\nabla^i(\dot\psi\overline\nabla_i\psi)dV=\int_S\dot\psi\overline\nabla_i\psi n^idS.
\end{align}
The momentum density $p_k= - \dot\psi\overline\nabla_k\psi$  of the scalar field obeys a conservation law as well:
\begin{align}
\label{36}
&\frac{d}{dt}\int_Vp_kdV = \frac{d}{dt}\int_V -\dot\psi\overline\nabla_k\psi dV
\\\nonumber
&= - \int_V\overline\nabla_i\Big{\{}\overline\nabla^i\psi\overline\nabla_k\psi 
- {\textstyle{\frac{1}{2} }}\delta^i_k \left[(\overline\nabla_j\psi)(\overline\nabla^j\psi) -\dot\psi^2\right]\Big{\}}dV\nonumber
\\\nonumber
&= -\int_ST^i_kn_idS.
\end{align}

Similarly, the angular~momentum  in the field obeys the conservation law
\begin{equation}
\frac{d}{dt}\int_V\varepsilon_{ijk}x^jp^kdV= - \int_S\varepsilon_{ijk}x^jT^k_sn^sdS.
\label{37}
\end{equation}
The angular~momentum  flux is given by the components $\varepsilon_{ijk}x^jT^{ks}$ through the surface $S$. The angular~momentum  density along the $z$-axis is
\begin{equation}
j_S= - \dot\psi\partial_\varphi \psi.
\label{s3eq9}
\end{equation}

Let us calculate the energy  density and the angular~momentum  density for rotating scalar waves. Consider a single $(l,m)$ component, like in (\ref{s2eq2}), written now in the form
\begin{equation}
\psi_{lm}={{\frac{1}{2} }}N_l^m\left[  Q_{lm}(t,r)e^{im\varphi }+ \overline Q_{lm}(t,r)e^{-im\varphi } \right] P^m_l(\mu)~,
\label{s3eq10}
\end{equation}
with $\mu=\cos\theta$, $P_l^m(\mu)$ are real Legendre functions. 
The energy density (\ref{s3eq5}) of the scalar field becomes 
 \begin{align}
\label{s3eq11}
&\varepsilon_S=\frac{1}{4} (N_l^m)^2 \Bigg\{ \left(  \dot Q_{lm}\dot{\overline Q}_{lm} +  Q'_{lm} \overline Q'_{lm}     \right)(P^m_l)^2
\\\nonumber
&+\frac{Q_{lm} \overline Q_{lm}}{r^2}\left[   \frac{m^2}{1-\mu^2} (P^m_l)^2+ (1-\mu^2)({P^m_l}')^2 \right]\nonumber
\\\nonumber
&+\frac{1}{2}\Big[   \big( \dot Q_{lm}^2  +{Q'}^2_{lm}  \big)  e^{2im\varphi }
+\big(  \dot{\overline Q}_{lm}^2 +\overline{Q'}^2_{lm}  \big)  e^{-2im\varphi }  \Big] (P^m_l)^2\nonumber
\\\nonumber
&+\frac{1}{2r^2}\left(  Q^2_{lm}e^{2im\varphi } +\overline Q^2_{lm}e^{-2im\varphi }  \right) \times
\\\nonumber
&~~~~~~~~~~~~~~~~~~~\left[  - \frac{m^2}{1-\mu^2}  (P^m_l)^2+\textstyle{(1-\mu^2)} ({P^m_l}')^2 \right] \Bigg\},
 \end{align}
 where we write ${P^m_l}'$ for $(d/d\mu) P^m_l$. If we average over $\varphi $ terms, we obtain
 \begin{align}
&\langle \varepsilon_S\rangle = \frac{1}{2\pi}\int_0^{2\pi}\!\!\!\!\!\varepsilon_Sd\varphi 
\\\nonumber
&~=\frac{1}{4} (N_l^m)^2 \Bigg\{ (\dot Q_{lm}\dot{\overline Q}_{lm} + Q'_{lm} \overline Q'_{lm})(P^m_l)^2
\\\nonumber
&+\frac{Q_{lm} \overline Q_{lm}}{r^2}
\left[  \frac{m^2}{1-\mu^2} (P^m_l)^2+\textstyle{(1-\mu^2)}({P^m_l}')^2 \right] \Bigg\}.
 \label{s3eq12}
 \end{align}

 If, in addition, we average over the latitude $\theta$ by taking ${\textstyle{\frac{1}{2} }\!}\int_{-1}^1\langle\epsilon_S\rangle d\mu$,  and use the integrals of $P'$'s evaluated in Appendix A, we arrive at the energy density,  averaged in every direction, as a function of $r$ in the form
 \begin{align}
\langle\!\langle\epsilon_S\rangle\!\rangle  = \frac{1}{16\pi} \left[
\dot Q_{lm} \dot{\overline Q}_{lm} +Q_{lm}' {\overline Q}_{lm}' +l (l+1) \frac{Q_{lm} \overline Q_{lm}}{r^2}
\right].
 \end{align}

The $z$-component of the angular momentum density is still simpler:
 \begin{align}
 j_S=\frac{im}{4}  (N_l^m)^2 \Big(\dot Q_{lm}\overline Q_{lm} - \dot{Q}_{lm} Q_{lm}e^{2im\varphi } \!- {\rm c.c.} \Big)(P^m_l)^2\!,
 \label{313}
 \end{align}
``c.c.'' denotes ``complex conjugation''.
 
So, after averaging over $\varphi $ and over $\theta$ (see Appendix A)  we get
\begin{equation}
\langle\!\langle j_S\rangle\!\rangle =
\frac{im}{16\pi}
\left(  \dot Q_{lm}\overline Q_{lm} - Q_{lm}\dot{\overline Q}_{lm}   \right)\!.
\label{314}
\end{equation}
\section{ Rotating linearized gravitational waves}
We consider first the linearized theory of gravity in a flat background  in spherical coordinates $(t,r,\theta,\varphi )$. It is natural to decompose the metric   perturbations  into tensor harmonics. These are given in terms of standard spherical harmonics (\ref{s2eq3}) and their derivatives, see, e.g., \cite{RW}, \cite{Ze}, \cite{NI}. For any symmetric second-rank covariant tensor $S_{\mu\nu}$ we define the expansion
\begin{align}
\label{s4eq1}
  &S_{\mu\nu}={\sum_{l,m}}\Big[  {\cal A}_{0lm}(t,r)a_{0lm\;\mu\nu} + {\cal A}_{1lm}(t,r)a_{1lm\;\mu\nu} 
\\\nonumber   &+  {\cal A}_{lm}(t,r)a_{lm\;\mu\nu} +  {\cal B}_{0lm}(t,r)b_{0lm\;\mu\nu} +  {\cal B}_{lm}(t,r)b_{lm\;\mu\nu} 
\\\nonumber   & + {\cal Q }_{0lm} (t,r)c_{0lm\,\mu\nu} +  {\cal Q }_{lm} (t,r)c_{lm\,\mu\nu}+{\cal D}_{lm}(t,r)d_{lm\,\mu\nu}  
\\\nonumber   &+ {\cal G}_{lm}(t,r)g_{lm\;\mu\nu}+{\cal F}_{lm}(t,r)f_{lm\;\mu\nu}           \Big],
\end{align}
where $a_{0lm\mu\nu}(\theta, \varphi ),\, ...,\,f_{lm\mu\nu} (\theta, \varphi )$ are tensor harmonics; we follow the notation and normalizations used by \cite{NI}. The harmonics can be classified into even- and odd-parities -- those with even-parity transform by the parity $(-1)^l$ under the inversion, i.e., the transformation $(\theta, \varphi ) ~\rightarrow~(\pi - \theta, \pi+\varphi )$, in the same way as $Y_{lm}(\theta, \varphi )$ -- while odd-parity harmonics are defined by the parity $(-1)^{l+1}$. The harmonics are mutually orthogonal and normalized over the sphere of radius $r$ in the following sense:
\begin{equation}
\int_0^{2\pi}\!\!\!\!\int_0^\pi\!\! \overline X_{l'm'\mu\nu}\delta^{\mu\rho}\delta^{\nu\sigma}X_{lm\rho\sigma}\sin\theta \,d\theta d\varphi=\delta_{l'l}\delta_{m'm},
\label{42}
\end{equation}
where
\begin{equation}
\delta^{\nu\sigma}={\rm diag}\{1,1,1/r^2,1/(r\sin\theta)^2\},
\end{equation}
 $X$ stands for any of the harmonics $a, ... , f$. The orthogonality properties enable us to find the ``coefficients" ${\cal A}_{0lm}$, etc. in the decomposition (\ref{s4eq1}) of the general covariant tensor $S_{\mu\nu}$. For example, in the following section we shall need the coefficient ${\cal Q }_{0lm}(t,r)$ multiplying $c_{0lm}$. 
Using the orthogonality property of the tensor harmonics and multiplying $S_{\mu\nu}$ by $\overline c_{0lm}$, we find
 \begin{equation}
 {\cal Q }_{0lm}(t,r)=\int_0^{2\pi}\!\!\!\!\int_0^\pi\delta^{\mu\rho}\delta^{\nu\sigma}S_{\mu\nu}\overline c_{0lm\,\rho\sigma}\sin\theta d\theta d\varphi .
 \label{s4eq4}
 \end{equation}
It is well known that the resulting field equations  for the radial factors may be discussed separately for odd- and even-parity perturbations of the Schwarzschild  black hole, see, for instance, \cite{RW}, \cite{GNPP}, \cite{Bi}, \cite{NI}, which of course remains true for flat space. For our purpose it is sufficient to consider odd-parity waves. In the expansions (\ref{s4eq1}) the odd-parity harmonics are $c_{0lm}, c_{lm}$ and $d_{lm}$, with explicit forms 
of the non-zero components
being
\begin{align}
\label{s4eq5}
c_{0lm\;t\theta} &= \frac{r}{\sqrt{2l(l+1)}}\frac{1}{\sin\theta}\partial_\varphi  Y_{lm},
\\
\label{s4eq6}
c_{0lm\;t\varphi} &= -\frac{r}{\sqrt{2l(l+1)}}  \sin\theta\partial_\theta Y_{lm},
\\
c_{lm\;r\theta} &= \frac{ir}{\sqrt{2l(l+1)}}  \frac{1}{\sin\theta}\partial_\varphi  Y_{lm},
\\
c_{lm\;r\phi} &= - \frac{ir}{\sqrt{2l(l+1)}}   \sin\theta\partial_\theta Y_{lm},
\\
d_{lm\;\theta\theta} &= \frac{ir^2}{\sqrt{2l(l+1)(l-1)(l+2)}}  \frac{1}{\sin\theta}X_{lm}, 
\end{align}
\begin{align}
d_{lm\;\varphi\varphi} &= \frac{-ir^2}{\sqrt{2l(l+1)(l-1)(l+2)}}  \sin\theta X_{lm}, 
 \\
\label{s4eq11}
 d_{lm\;\theta\varphi} &= \frac{-ir^2}{\sqrt{2l(l+1)(l-1)(l+2)}}  \sin\theta W_{lm},
\end{align}
where
\begin{align}
\label{s4eq12}
X_{lm}&=2\partial_\varphi \left(\partial_\theta - \cot\theta\right) Y_{lm}~,
\\\nonumber
W_{lm}&=\left(\partial^2_\theta  - \cot\theta\partial_\theta - \frac{1}{\sin^2\theta}\partial^2_\varphi    \right) Y_{lm}.
\end{align}
For the explicit forms of the other tensor harmonics, see \cite{RW}, \cite{Ze}, \cite{Bi} and \cite{NI}. As mentioned above, our convention for harmonics is  that used in \cite{NI} which differs slightly from \cite{RW} or \cite{Ze}. However, following \cite{NI} we assume the expansion of the odd-parity metric perturbations  $h^{(i)}_{\mu\nu}$ ($(i)=(1),(2)$ denotes the first- and second-order perturbations) to have the form
\begin{align}
\label{s4eq13}
\nonumber
h^{(i)}_{\mu\nu}=\sum_{lm}\Bigg[&-\frac{\sqrt{2l(l+1)}}{r}  h^{(i)}_{0lm}(t,r)c_{0lm\,\mu\nu} 
\\
&+ i \frac{\sqrt{2l(l+1)}}{r}  h^{(i)}_{1lm}(t,r)c_{lm\,\mu\nu}
\\\nonumber
&+i \frac{\sqrt{2l(l+1)(l-1)(l+2)}}{2r^2} h^{(i)}_{2lm}(t,r)d_{lm\,\mu\nu} \Bigg],
\end{align}
so that the factors multiplying the harmonics (\ref{s4eq5}-\ref{s4eq11}) cancel the multiplicative factors at $h^{(i)}_{0lm }, h^{(i)}_{1lm }, h^{(i)}_{2lm }$ in (\ref{s4eq13}). Hence the radial functions, the $h$'s, are identical in \cite{RW}, \cite{Ze}, \cite{Bi} and \cite{NI}.

In the literature time-dependent perturbations  are usually Fourier analyzed and perturbations  with even parity discussed. In
\cite{Bi} a complete account of the Hamiltonian, Regge-Wheeler-Zerilli and Newman-Penrose formalism for perturbations (with both parities) of a Reissner-Nordstr\"om black hole is given and the dipole $l=1$ perturbations  are also considered. Putting $M=e=0$ there and choosing the Regge-Wheeler gauge condition for odd-parity perturbations, $h^{(i)}_{2lm\mu\nu}=0$, we obtain the radial functions $ h^{(i)}_{0lm\mu\nu}(t,r)$ and $ h^{(i)}_{1lm\mu\nu}(t,r)$ in terms of a function $R$ satisfying the wave equation\footnote{One can get these equations  from the original paper \cite{RW} as well as after going back from their Fourier decompositions in $e^{-ikt}$ to the time-dependence and after correcting several misprints in their equations (22).} as follows (discarding labels):
\begin{align}
&h_0 = - \frac{1}{(l-1)(l+2)}(rR)'~,
\label{s4eq14}
\\
&h_1 =- \frac{r}{(l-1)(l+2)}\dot R~,
\\
&- \ddot R+R''-\frac{l(l+1)}{r^2}R=0.
\end{align}
We see that last equation can be put in the form of Eq. (\ref{s2eq4}) for the scalar field amplitude $Q_l(t,r)$ through the substitution 
\begin{equation}
R=rq.
\label{48}
\end{equation}
So in terms of $q(t,r)$ the odd-parity metric perturbations are given by
\begin{align}
h_0&= - \frac{1}{(l-1)(l+2)}(r^2q)',
\label{s4eq18}
\\
h_1&= - \frac{1}{(l-1)(l+2)}r^2\dot q.
\label{s4eq19}
\end{align}
These formulas are valid for $l\ge 2$; in the linear theory in vacuum the dipole odd-parity perturbations ($l=1$) can be transformed away by a simple gauge transformation  \cite{Ze,Bi}. 
Later, in Section VI, we shall substitute for $q(t,r)$ the solution given in (\ref{s2eq8}) and consider (\ref{s4eq18}-\ref{s4eq19})
to be the first-order perturbations  $h^{(1)}$ generating the second-order perturbations.


\section{ Energy in gravitational waves }
 In \cite{KLB06} we defined gravitational energy  on $x^0$=const slices in stationary asymptotically  flat spacetimes  by taking the difference between the total mass-energy and the total ``mechanical" energy arising from the matter stress tensor. Many formulas given in \cite{KLB06} are true in general, non-stationary spacetimes $\!\!$.

Consider the slice $x^0$=const and the metric in the $1+3$ decomposition associated with the Hamiltonian formulation (see, e.g., \cite{MTW})
\begin{align}
ds^2&=g_{00}dt^2+2g_{0k}dtdx^k+g_{ik}dx^idx^k
\label{s5eq1}
\\\nonumber
&=(w^2-W^2)dt^2+2W_kdtdx^k-\gamma_{ik}dx^idx^k,
\end{align}
 $i, k=1, 2, 3$, $w=\sqrt{g_{00}+W^2}$ and $W^2=\gamma_{ik}W^iW^k$ (cf. (2.36) in \cite{KLB06}). The energy of matter can be calculated using the Einstein constraint equation  for
 \begin{equation}
 2wG^{00}={\cal R} + K^2 - K^{ik}K_{ik},
 \label{52}
 \end{equation}
 where ${\cal R}$ is the scalar curvature of the 3-metric $\gamma_{ik}$ and the external curvature of $x^0$=const slices is
 \begin{equation}
K_{ik}=w^{-1}\left(\nabla_{(i}W_{k)}   -{\textstyle{\frac{1}{2} }\!} \partial_t\gamma_{ik}   \right)~~,~~K=\gamma^{ik}K_{ik},
\label{s5eq3}
 \end{equation}
 $\nabla_k$ is a $\gamma$-covariant derivative. The total (gravitational  and matter) mass-energy as measured at infinity is $M$. The gravitational  field energy on the $x^0$=const slice can be defined as
 \begin{equation}
 E_G=M -\frac{1}{2\kappa}\int({\cal R}+K^2-K^{ik}K_{ik})\sqrt{\gamma}d^3x~,
 \label{s5eq4}
 \end{equation}
$\kappa$ is Einstein's gravitational constant, $\gamma=\det\gamma_{ik}$, all volume integrals here  and below are taken over $x^0$=const. 

It is often advantageous to use more general spatial coordinates  than Cartesian coordinates  at infinity, so assume that the metric of the background  Minkowski  spacetime  which at infinity is ``the limiting spacetime" to the physical spacetime, has the general form\footnote{Details on how suitable background metrics and mappings between backgrounds  and physical spacetime  may be introduced can be found in \cite{BK05}, \cite{KBL97}.}:
\begin{equation}
 d\overline s^2= dt^2 - \overline\gamma_{ik}dx^idx^k.
 \label{55}
\end{equation}
Denoting $\overline\gamma$-covariant derivatives by $\overline\nabla_k$ and introducing tensorial quantities
\begin{align}
\Delta^m_{~ik}&=\Gamma^m_{~ik} - {\overline\Gamma}^m_{~ik} 
\\\nonumber
&={\textstyle{\frac{1}{2} }\!}\gamma^{mn}\left( \overline\nabla_i\gamma_{nk} +\overline\nabla_k\gamma_{ni} - \overline\nabla_n\gamma_{ik}\right)\!,
\label{56}
\end{align}
one can write the scalar 3-curvature density entering (\ref{s5eq4}) as follows:
\begin{equation}
\frac{1}{2\kappa}\sqrt{\gamma}{\cal R}=\frac{1}{2\kappa}\sqrt{\gamma}{\cal L} - \frac{1}{2\kappa}\partial_i(\sqrt{\gamma}k^i).
\label{s5eq7}
\end{equation}
Here ${\cal L}$ is the Einstein Lagrangian
\begin{equation}
{\cal L}=-\gamma^{ik}\left(  \Delta^m_{~ik}\Delta^n_{~mn} - \Delta^m_{~in}\Delta^n_{~mk} \right),
\label{s5eq8}
\end{equation}
and $k^i$ are given by (cf. (3.16) in  \cite{KLB06}):
\begin{equation}
k^i=\gamma^{-1}\overline\nabla_k(\gamma\gamma^{ik}).
\label{s5eq9}
\end{equation}
In an asymptotically  flat spacetime  the integral of the second term on the right-hand side of (\ref{s5eq7}) gives again the mass,
\begin{equation}
- \frac{1}{2\kappa}\int\partial_i(\sqrt{\gamma}k^i)d^3x= - \frac{1}{2\kappa}\oint \sqrt{\gamma}k^i dS_i=M.
\label{s5eq10}
\end{equation}
This is the total mass of spacetime  as measured at spatial infinity. In our case of pure gravitational waves  the surface integral gives the total mass-energy of 
the waves. In a general case of radiation going through spatial infinity for all times, time-dependent radiative terms $\thicksim 1/r$ would also contribute to the surface integral (\ref{s5eq10}), however for pulse-type radiation considered in this work only the {\it static} mass-energy results. Returning to $E_G$ in (\ref{s5eq4}), substituting from (\ref{s5eq7}) and (\ref{s5eq10}) we find the gravitational energy as a volume integral over the slice $x^0$=const:
\begin{align}
\label{s5eq11}
E_G&=\frac{1}{2\kappa}\int\left( K^{ik}K_{ik} - K^2-{\cal L}\right)\sqrt{\gamma}\;d^3x
\\\nonumber
&= \int\varepsilon_G\sqrt{\gamma}\;d^3x,
\end{align}
with ${\cal L}$ and $K_{ik}$ given in (\ref{s5eq8}) and (\ref{s5eq3}). The energy density $\varepsilon_G$ is a scalar density on $x^0$=const due to the use of the background metric\footnote{It can be shown that (\ref{s5eq11}) is gauge invariant provided that the rules of the Brill-Hartle averaging employed in the classical work of Isaacson \cite{Is68-2} are used (for example, under integrals divergences become small);  in the TT-gauge $\langle \varepsilon_G\rangle $ becomes $\thicksim \dot h_{ik}\dot h^{ik}$ of Isaacson.}.

Assume now the physical metric $\gamma_{ik}$ to be close to the background  metric $\overline\gamma_{ik}$ 
and $g_{00}=1+h_{00}$ so that
\begin{equation}
\gamma_{ik}=\overline\gamma_{ik}+h_{ik}~~~,~~~W_k=h_{0k}~~~,~~~w=1+\frac{1}{2}h_{00}.
\end{equation}
The terms proportional to ${\cal O}(h^2)$ will be neglected. The extrinsic curvature (\ref{s5eq3}) then becomes
\begin{equation}
K_{ik}=\overline\nabla_{(i}h_{k)0} - {{\frac{1}{2} }}\dot h_{ik},
\label{s5eq13}
\end{equation}
the Lagrangian (\ref{s5eq9}) is
\begin{align}
\label{s5eq14} 
- {\cal L}= &{\textstyle{\frac{1}{2} }\!}\overline\nabla_mh\overline\nabla_kh^{km} - \textstyle{\frac{1}{4}}\overline\nabla_mh\overline\nabla^mh
\\\nonumber
& -{\textstyle{\frac{1}{2} }\!} \overline\nabla^i h^{km}\overline\nabla_mh_{ik}+ \textstyle{\frac{1}{4}} \overline\nabla^i h^{km}\overline\nabla_ih_{km}.
\end{align}

Since we are considering waves by using tensor spherical harmonics it is natural to choose the background  metric $\overline\gamma_{ik}$ in spherical coordinates:
\begin{equation}
\overline\gamma_{11}=1~~~,~~~\overline\gamma_{22}= r^2~~~,~~~\overline\gamma_{33}=r^2\sin^2\theta.
\end{equation}
We start from the metric for odd-parity perturbations  in the general form (\ref{s4eq13}) with harmonics given by (\ref{s4eq5}-\ref{s4eq12}). However, since in the odd-parity case only four (non-diagonal) components are non-vanishing, the following simple notation will turn out to be useful:
\begin{align}
\nonumber
h_{02}&=h_0f_2~,~h_{03}=h_0f_3~,~h_{12}=h_1f_2~,~h_{13}=h_1f_3~,
\\
f_2 & = -\frac{1}{\sin\theta}\partial_\varphi  Y_{lm}~,~f_3=\sin\theta\partial_\theta Y_{lm}.
\label{s5eq16}
\end{align}
Starting from these expressions, calculating $K_{ik}$ and ${\cal L}$ from (\ref{s5eq13}), (\ref{s5eq14}) using $\overline\nabla_k$ in flat-space spherical coordinates, we arrive, after straightforward - though not short - calculations at the following expression for the gravitational energy  density of odd-parity waves
\begin{equation}
\varepsilon_G=\frac{1}{2\kappa}\left( K^{ik}K_{ik}-K^2 - {\cal L}\right)\!,
\end{equation}
where
\begin{align}
\nonumber
K^{ik}K_{ik}&-K^2=
\frac{h_0^2}{2r^4\sin^2\theta}\Big[ \left(    \partial_\varphi  f_2+\partial_\theta f_3-2\cot\theta f_3   \right)^2 
\\
&+\frac{1}{2r^4}  \left( r\dot h_1 - rh_0' +2h_0\right)^2 \left( f_2^2+\frac{1}{\sin^2\theta}f_3^2\right)\nonumber
\\
& - 4\partial_\theta f_2\left(   \sin\theta\cos\theta f_2 +\partial_\varphi  f_3  \right)\Big]\!,
\label{s5eq18}
\end{align}
and
\begin{align}
\nonumber
-{\cal L}=\frac{1}{2r^4}\Big[& 4h_1(rh'_1 - h_1)   \left( f_2^2+\frac{1}{\sin^2\theta}f_3^2\right)   
\\
&  +\frac{h_1^2}{\sin^2\theta}\left(  \partial_\varphi  f_2 - \partial_\theta f_3     \right)^2 \Big]\!.
\label{s5eq19}
\end{align}

 We now express the radial parts of perturbations  from (\ref{s4eq18}-\ref{s4eq19}) in terms of the gauge invariant quantity $q$ satisfying the wave equation  (\ref{s2eq4})  ($Q$ there is denoted by $q$ in the gravitational case), and functions $f_2, f_3$ and their derivatives in terms of spherical harmonics according to (\ref{s5eq16}). The individual terms in (\ref{s5eq18}) to (\ref{s5eq19}) are processed as follows. Each term is quadratic in ``$hf$", say $h_0 f_2 h_1 f_3$. To get real results, we should take\footnote{\label{foononcom}Notice that the terms cannot be commuted since, for example, $Re(h_0f_2)Re(h_1f_3)\ne Re(h_0f_3)Re(h_1f_2)$ so that when evaluating (\ref{s5eq18}), (\ref{s5eq19}) the {\it corresponding terms} must be grouped together.}
\begin{align}
\nonumber
&Re(h_0f_2) Re(h_1f_3)={\textstyle{\frac{1}{2} }\!}(h_0f_2+\overline h_0\overline f_2) {\textstyle{\frac{1}{2} }\!}(h_1f_3+\overline h_1 f_3)
\\
&=\textstyle{\frac{1}{4}}(h_0f_2\overline h_1\overline  f_3+\overline h_0\overline f_2h_1f_3)+{\rm terms ~in}~e^{\pm2im\varphi }.
 \label{519}
\end{align}
The last terms drop out after averaging over $\varphi $ -- see the scalar-field case, equation   (\ref{s3eq11}) -- so we retain only the first terms with ($\textstyle{\frac{1}{4}}$) expressed in terms of $q_{lm}(t,r)$ and $P^m_l(\mu)$. In this way we obtain, after somewhat lengthy but straightforward calculations and arrangements, the $\varphi $-averaged energy density of odd-parity rotating gravitational waves:
\begin{align}
\nonumber
& 2\kappa\langle\varepsilon_G\rangle ={\frac{1}{4}}[l(l+1)]^2{\Bigg\{}  (\dot q\dot{\overline q}+q'\overline q' )P^2 
\\\nonumber
+&\frac{q\overline q}{r^2} \left[    \frac{m^2}{1-\mu^2}P^2   +{(1-\mu^2)}{P'}^2     \right]                {\Bigg\}}
+\frac{[l(l+1)]^2}{2r^3}  ( r^2q\overline q)'    P^2 
\\\nonumber
&+\frac{1}{2r}(r^2\dot q\dot{\overline q})' \left[   { \frac{m^2}{1-\mu^2}}P^2   +\textstyle{(1-\mu^2)}{P'}^2     \right]
\\ \nonumber 
&+\left[ q'{\overline q}' +\frac{2}{r}(q{\overline q})' \right]\Bigg[  { \frac{m^2}{1-\mu^2}\left(    \frac{m^2+\mu^2}{1 - \mu^2} -l(l+1)           \right)  }P^2
\\
\nonumber
&+ \left({ \frac{4m^2}{1-\mu^2} }- l(l+1)        \right)\mu PP' +\textstyle{(\mu^2+m^2)}{P'}^2    \Bigg]
\\ \nonumber 
&+\frac{q\overline q}{r^2}\Bigg{\{}\!{\frac{m^2}{1-\mu^2}}\!\left[  \!1\! - \!5l(l+1)\!+ \! { \frac{4(m^2+\mu^2)}{1-\mu^2}  }  \!\right] \!P^2
\\ \nonumber
&+\left(  \!{\frac{4m^2}{1-\mu^2}} \!-\textstyle{l(l+1)} \! \right)\!\mu PP'
\\ 
&+\left[{1\!+\!m^2-(1\!-\!\mu^2)l(l+1) }\right] {P'}^2 \Bigg{\}}. ~~~~~~~~~~
\end{align}
Here $P$ and $P'$ denote $P^m_l(\mu)$ and $d_\mu P^m_l(\mu)$; $q$ stands for $q_l(t,r)$.

This result has some noteworthy properties. The first terms in the first curly brackets coincide precisely with the energy (\ref{s3eq12})  in scalar waves averaged over $\varphi $. Together with the next term in $[l(l+1)]^2$ they dominate for big $l$. Nevertheless, for rotating waves when $m\thicksim l$ the scalar field part dominates. The very first term $\thicksim [l(l+1)]^2\dot q\dot{\overline q}$ will dominate for high frequency waves $(\dot q\dot{\overline q} \thicksim \omega^2)$ and high $l$ for both scalar and gravitational waves.

\section{ Second-order odd parity dipole perturbations}
In general the second-order metric perturbations  $h^{(2)}$  can be obtained by solving the equations 
\begin{equation}
G_{\mu\nu}^{(1)}[{h^{(2)}}]= - G_{\mu\nu}^{(2)}[h^{(1)} ,h^{(1)}],
\label{s6eq1}
\end{equation}
where $h^{(i)}$ represent quantities given in (\ref{s4eq13}) with all indices seen explicitly.
The right-hand side   of (\ref{s6eq1}) is the source term in the form of an effective energy-momentum  tensor quadratic in the first-order perturbations  $h^{(1)}$ and their derivatives known from the solutions of $G_{\mu\nu}^{(1)}[h^{(1)}]=0$. In principle 
writing down the right-hand side of Eq. (\ref{s6eq1}) and solving it is a clear task, in practise a formidable one.

However, to determine the influence of gravitational waves  on the rotation of local inertial frames  at the axis of symmetry we do not need to solve (\ref{s6eq1}) in general. First, following our method in the translational symmetric case \cite{BKL08cqg1}, \cite{BKL08cqg2}, we shall average over $\varphi $ so that the source term on the right-hand side  will become axisymmetric. Before averaging it will contain terms of the form $\partial_th^{(1)}\partial_\varphi  h^{(1)}$ each of which involves two terms independent of $\varphi $ and two terms varying like $\sin 2m\varphi  $ or $\cos 2m\varphi $. Those terms cannot cause any rotation of the inertial frames  on the axis. Hence, we shall concentrate on the terms that are independent of $\varphi $ and denote them again, as in previous sections, by the average symbol $\langle \rangle $. In addition, as in the translational symmetric case the dragging will be fully determined by solutions of just one equation  in (\ref{s6eq1}) -- that with $\mu=t, \nu=\varphi $.

Therefore, we consider the equation 
\begin{equation}
G_{t\varphi }^{(1)}[h^{(2)}] = - \langle G_{t\varphi }^{(2)}[h^{(1)} ,h^{(1)}]\rangle,
\label{s6eq2}
\end{equation}
or equivalently,
\begin{equation}
R_{t\varphi }^{(1)}[h^{(2)}] = - \langle R_{t\varphi }^{(2)}[h^{(1)} ,h^{(1)}]\rangle,
\label{s6eq3}
\end{equation}
which follows from (\ref{s6eq2}) because in the background $\overline g_{\mu\nu}$ is the flat metric in spherical coordinates and $R_{\mu\nu}^{(1)}[h^{(1)}] =0$.
To solve these equations  we shall expand both sides in tensor spherical harmonics. Any symmetric second-rank covariant tensor can be expanded as in (\ref{s4eq1}). We are interested in the axisymmetric $(t,\varphi )$ component, so the only harmonic entering the left-hand side  of (\ref{s6eq3}) linear in $h^{(2)}$ is the odd-parity harmonic $c_{0lm}$ given in (\ref{s4eq6}).
In order to extract the contribution proportional to the same harmonic on the right-hand side  of (\ref{s6eq3})
 we imagine it to be expanded as the general covariant tensor $S_{\mu\nu}$ in (\ref{s4eq1}).
The coefficient ${\cal Q }_{0lm}(t,r)$ multiplying $c_{0lm\,\mu\nu}$ can be derived using formula (\ref{s4eq4}). In our case of the averaged  -- and thus axially symmetric -- component ~~~~~~~~~~$\langle R_{t\varphi }^{(2)}[h^{(1)},h^{(1)}]\rangle $ we put $m=0$ and $(\mu\nu)=(t\varphi )$ to obtain
\begin{align}
{\cal Q }_{0l0}(t,r)&= \int_0^{2\pi}\!\!\! \int_0^\pi -2\langle R_{t\varphi }^{(2)}\rangle \overline c_{0l0\;t\varphi } \frac{\sin\theta d\theta}{r^2\sin^2\theta} ~ d\varphi
\nonumber\\
&=\frac{4\pi}{\sqrt{2l(l+1)}r}\int_0^\pi\langle R_{t\varphi}^{(2)}\rangle \partial_\theta Y_{l0}\;d\theta ,
\label{65}
\end{align}
where we substituted for $\overline c_{0l0\;t\varphi }$ from (\ref{s4eq6}). Hence, the right-hand side  of equation  (\ref{s6eq3}) reads
\begin{equation}
2\pi\Bigg{\{}\!\!\int_0^\pi\!\!\!\!\langle R_{t\varphi }^{(2)}[h^{(1)}\!,h^{(1)}]\rangle \partial_\theta Y_{l0}d\theta\Bigg{\}}\left(   -\frac{\sin\theta\partial_\theta Y_{l0}}{l(l+1)}\right).
\label{s6eq6}
\end{equation}

The left-hand side  of (\ref{s6eq1}) in flat background  (in spherical coordinates) is formed from the second derivatives of the second-order perturbations:
\begin{align}
G_{\mu\nu}^{(1)}[h^{(2)}]= -{{\frac{1}{2} }\!} \Big[ &
  {h^{(2);\alpha}_{\mu\nu;\alpha}}
+ {h^{(2);\alpha}_{\mu\alpha;\nu}}
+ {h^{(2);\alpha}_{\nu\alpha;\mu}}  - h^{(2)\alpha}_{~\;\alpha\;;\mu\nu}
\nonumber\\
 & -\overline g_{\mu\nu} \!\left(   h^{(2)~;\alpha\beta}_{~\beta\alpha}  -  h_{~\beta~~~;\alpha}^{(2)\beta;\alpha}\right)   \Big].
\label{s6eq7}
\end{align}
We expand $G_{\mu\nu}^{(1)}$ in tensor spherical harmonics  and realize that we need only the axisymmetric component $G_{t\varphi }^{(1)}$; the result will thus be proportional to $c_{0l0}$. It is now important  that by using the  gauge transformation $x'\rightarrow x+\xi$ to the second order we can choose the Regge-Wheeler gauge also for the second-order perturbations  \cite{GNPP}, \cite{NI}, so for odd-parity perturbations  we can again set $h^{\,(2)}_2=0$ and keep just $h^{\,(2)}_0$  and $h^{\,(2)}_1$.

Calculations of $G_{t\varphi }^{(1)}[h^{(2)}]$ from (\ref{s6eq7}) lead us to the following expression:
\begin{equation}
 {{\frac{1}{2} }\!} \left[
 -{h_0^{(2)}}''\!\! +\frac{l(l+1)}{r^2}h^{(2)}_0\!
+{{\dot h}_1^{(2)}}{}'\!+\!\frac{2}{r}{\dot h^{(2)}}_1 \right]\sin\theta\partial_\theta Y_{l0}.
\end{equation}

Combining this with the right-hand side (\ref{s6eq6}) we arrive at the differential equation  implied by the field equation  (\ref{s6eq2})
for radial parts in the form
\begin{align}
\label{s6eq9}
&{{\frac{1}{2} }\!}\left[   {h_0^{(2)}}''\!\!\! - \frac{l(l+1)}{r^2} {h_0^{(2)}} -{{\dot h}_1^{(2)}}{}'\!-\frac{2}{r} {{\dot h}_1^{(2)}} \right]
\\
\nonumber
&~~~~~~~~~~~~~~~~ =\frac{2\pi}{l(l+1)}\int_0^\pi\!\!\!\!  \langle R_{t\varphi }^{\,(2)}[h^{(1)} ,h^{(1)}]\rangle \partial_\theta Y_{l0}  \,\, d\theta.
\end{align}
This equation  suggests that we need to consider more field equations  to determine both second-order perturbations  
$h^{(2)}_0$  and $h^{(2)}_1\!\!$ from the first-order ones. Indeed, if we wish to calculate both
$h^{(2)}_0$  and $h^{(2)}_1$ for all $l$'s we have to take into account equations  (\ref{s6eq1}) also for $\textstyle{(\mu\nu)=(\theta,\varphi ),(r,\varphi )}$ and by combining them we can achieve just one wave-type equation  for $h^{(2)}_1$:
\begin{equation}
r \left(\partial_{t}^{2}-\partial_{r}^{2}+\frac{l(l+1)}{r^2}\right) \frac{h^{(2)}_{1}}{r} =S_{r\varphi }+S_{\theta\varphi }' - \frac{2}{r}S_{\theta\varphi },
\label{610}
\end{equation}
where components $S_{\mu\nu}$ stand for  the averaged components $\langle R_{\mu\nu}^{(2)}[h^{(1)} ,h^{(1)}]\rangle $. This is a hyperbolic equation  indicating that for general $l$ the effects of the first-order terms are perturbations  of the second-order terms that are non-instantaneous. However, in all our previous work on dragging of inertial frames  due to angular~momentum  of the sources (see, for example, \cite{BKL07} and references therein), the effects were instantaneous, at least at this lowest order. In fact, the same situation arises here. Inertial frames at the origin will be influenced primarily by the {\it dipole} perturbations  since all waves behave as $r^l$ there. Now it is well known that for $l=1$ one can achieve $h_1=0$ by an appropriate gauge transformation \cite{RW}, \cite{Bi}. For the dipole second-order perturbations  the equation  (\ref{s6eq9})
becomes
\begin{equation}
{h^{(2)}_{\;0}}'' -  \frac{2}{r^2}  h^{(2)}_{\;0} = g(t,r) ,
\label{s6eq11}
\end{equation}
where
\begin{equation}
g(t,r) = - 
\sqrt{ \frac{3}{4\pi} } 
\int_0^\pi\!\!
\int_0^{2\pi}\!\!\!\!\!\!R_{t\varphi }^{(2)}[h^{(1)} ,h^{(1)}]
\sin\theta \,d\theta d\varphi.
\label{s6eq12}
\end{equation}
Here we indicated the $\varphi $-averaging explicitly (which cancelled the factor $2\pi$) and we substituted for $\partial_\theta Y_{10}$. With $g(t,r)$
known, the solution of (\ref{s6eq11}) can easily be found by variation of constants from the simple solutions proportional to $1/r$ and $r^2$ of the homogeneous equation. Putting the arbitrary constants entering the homogeneous solution equal to zero, the complete solution for $h^{(2)}_{\;0}(t,r)$ reads
\begin{equation}
h^{(2)}_{\;0}= - \frac{1}{3r}\int_0^r \!\!g(t,r')\,r'^2 dr' - \frac{r^2}{3}\int_r^\infty\!\! g(t,r')\,\frac{dr'}{r'}.
\label{s6eq13}
\end{equation}

\section{How rotating waves drag inertial frames}
Since we are interested in the dragging of inertial frames  near the origin we wish to express
\begin{equation}
h^{(2)}_{0}(t,r)\simeq - \frac{r^2}{3}\int_0^\infty g(t,r')\frac{1}{r'}dr',
\end{equation}
with $g(t,r)$ given by (\ref{s6eq12}) and $\langle R^{(2)}_{t\varphi}\rangle$ in Appendix B.

The general form of the odd-parity metric perturbations  (\ref{s4eq13}), with the tensor spherical harmonics  (\ref{s4eq12}), for the special case of $l=1,m=0$ perturbations  with only $h^{\!^{(2)}}_{0\,10}\ne0$ yields the metric component $g_{t\varphi }$ near the origin in the form
\begin{equation}
g_{t\varphi }^{(2)}= - \sqrt{\frac{3}{4\pi}} h^{(2)}_{0}(t,r)\sin^2\theta = - \omega_0 \, r^2\sin^2\theta,
\end{equation}
where the rotation of an inertial frame  (of a gyroscope) located near the origin has the angular velocity
\begin{equation}
\omega_0=  \frac{1}{4 \pi} \int_0^\infty  \int R_{t\varphi }^{(2)}[h^{(1)} ,h^{(1)}]~ d\Omega~ \frac{dr}{r},
\label{Omega0}
\end{equation}
where $d\Omega = \sin \theta\, d\theta\, d\varphi$.

To evaluate $R_{t\varphi }^{(2)}[h^{(1)} ,h^{(1)}]$, which has more complicated structure than any expression calculated
so far, we found it more convenient to start from the real form of the solution of the original scalar wave equation (\ref{s2eq1}) and
use it to express the gravitational wave perturbations -- see Eqs. (\ref{s7eq4}), (\ref{s7eq5}) below.
As a consequence, also the metric coefficients will be real from the start, rather than of the
form (\ref{s5eq16}). With real expressions one does not have to be careful when grouping together properly
the corresponding terms in the quadratic expressions like we had to do in calculating the energy
densities (cf. footnote 4) and it is much easier to 
find the results using the standard (commutative) computer algebra codes such as {\tt tensor} package in {Maple}. 
In fact, we first proceeded by using the same methods as in the case of energies and, after going to the real
variables, we checked that both results agree.

So now we start from the metric perturbations written as follows
\begin{align}
      h_{t \theta}&=\frac{1}{\sin \theta} \frac{\partial \chi}{\partial r \partial \varphi}
~,    &h_{t \varphi}&=-{\sin\theta} \frac{\partial \chi}{\partial r \partial \theta}~,
\label{s7eq4}
\\
      h_{r \theta}&=\frac{1}{\sin\theta} \frac{\partial \chi}{\partial t \partial \varphi}
~,    &h_{r \varphi}&=-{\sin\theta} \frac{\partial \chi}{\partial t \partial \theta}~,
\label{s7eq5}
\end{align}
where function $\chi(t,r,\theta,\varphi) = r^2 \psi_{lm}$, ($\psi$ satisfies the scalar wave equation (\ref{s2eq1}) in flat
space); introducing $\chi$ instead of $\psi$ turns out to simplify the second-order Ricci tensor. Using the
general result for the the component $R^{(2)}_{t\varphi}$ (see (\ref{appBeq2}) in Appendix B) and substituting for $\chi$ from
(\ref{s7eq4}-\ref{s7eq5}) we arrive at
\begin{align}
\label{s7eq6}
&R^{(2)}_{t\varphi} = -\frac{1}{2}\, \frac {1}{{r}^{2}}\Bigg( 
2\,\chi_{,\theta {r} {r} \theta}  \chi_{,\varphi t}-\chi_{,r \varphi r}  \chi_{,\theta t \theta}
\\\nonumber
&+2\,\chi_{,t r \varphi \theta}  \chi_{,r \theta}-\chi_{,{r} {r} \varphi \theta}  \chi_{,t \theta}-\chi_{,\theta t r \theta}  \chi_{,\varphi r}+\chi_{,{r} {r} \theta}  \chi_{,t \varphi \theta}
\\\nonumber
&
+\chi_{,\theta r \theta}  \chi_{,r \varphi t}
-\chi_{,t \theta}  \chi_{,{t} {t} \varphi \theta}-\chi_{,{t} {t} \theta}  \chi_{,t \varphi \theta}
-\chi_{,\varphi t}  \chi_{,\theta {t} {t} \theta} \Bigg)
\\\nonumber
&
+{\frac {\left (\chi_{,r \theta}  \chi_{,r \varphi t} -\chi_{,t r \theta}  \chi_{,\varphi r} -\chi_{,r \varphi r}  \chi_{,t \theta}
+\chi_{,{t} {t} \theta}  \chi_{,\varphi t}\right )\cos \theta}{2{r}^{2}\sin\theta}}
\\\nonumber
&-{\frac {\chi_{,r {\varphi} {\varphi} t}  \chi_{,\varphi r}
+\chi_{,{\varphi} {\varphi} r}  \chi_{,r \varphi t}-\chi_{,t \varphi t}  \chi_{,{\varphi} {\varphi} t} }{2{r}^{2} \sin^2\theta }}
+{\frac {\chi_{,r \theta} \chi_{,\varphi t}\cos\theta}{{r}^{3}\sin\theta}}
\\\nonumber
&
+{\frac {\chi_{,\theta r \theta}  \chi_{,\varphi t}}{{r}^{3}}}
+{\frac {\chi_{,\varphi t}}{2 {r}^{3}\sin^2\theta }}\left( 2\chi_{,{\varphi} {\varphi} r}+2r\chi_{,{\varphi} {\varphi} t t}  - r \chi_{,{\varphi} {\varphi} r r}\right)  ~.
\end{align}
We shall now write the real part of the scalar potential $\chi(t,r,\theta,\varphi)$ in the form corresponding to the form (\ref{s2eq13}) as the product
\begin{align}
\label{chi_fce}
 \chi
= \tilde B_l N^m_l\, \kappa(t,r) P_{l}^m(\cos\theta) \cos( m\varphi-\lambda(t,r) )~,
\end{align}
where $\tilde B_l=B_l\, 2^l l!$, $N^m_l$ and $\lambda$ are given in (\ref{s2eq14}) and (\ref{s2eq15}), and 
we have introduced function 
\begin{align}
 \label{s7eq8}
 \kappa(t,r) &=  \frac{\tilde r^{l+2} }{[ (1+\tilde r^2-\tilde t^2)^2 + 4 \tilde t^2 ]^{(l+1)/2} }~.
\end{align}
Inserting the function $\chi$ from (\ref{chi_fce}) into (\ref{s7eq6}) and employing the formulas for the integrals of the products
of the Legendre functions given in Appendix A to find the average over the spherical angles we first
find
\begin{align}
\label{s7eq9}
\nonumber
&\int R^{(2)}_{t \varphi}\, d\Omega = -
\tilde B_l^2 \frac{m}{r^2} \Big [
2\,\kappa\lambda_{{,r}}\kappa_{{,r t}}
-2\,{\kappa_{{,r}}}^{2}\lambda_{{,t}}
-2\,\kappa_{{,r}}\kappa\lambda_{{,r t}}
\\\nonumber
+&2\left ({l}^{2}+l+1\right )\kappa_{{,t}}\kappa_{{,r}}\lambda_{{,r}}
-\frac{2}{r}\,{{l\left (l+1\right )\kappa\left (\lambda_{{,r}}\kappa_{{,t}}-\kappa_{{,r}}\lambda_{{,t}}\right )}}
\\\nonumber
+&\left ({l}^{2}+l+2\right ) \kappa \left (\kappa_{{,t}}\lambda_{{,r r}}-\lambda_{{,t}}\kappa_{{,r r}}\right )
- l\left (l+1\right )\kappa^{2}{\lambda_{{,t}}}^{3}
\\
+&l\left (l+1\right )\left ( \kappa ^{2}\lambda_{{,t}}{\lambda_{{,r}}}^{2}
-2\,{\kappa_{{,t}}}^{2}\lambda_{{,t}}+\kappa\kappa_{{,t t}}\lambda_{
{t}}-\kappa\lambda_{{,t t}}\kappa_{{,t}}\right )\Big ].
\end{align}

\begin{figure}[b]
\includegraphics[angle = 270, width=8.4cm,clip=true,bb=65 104 546 744]{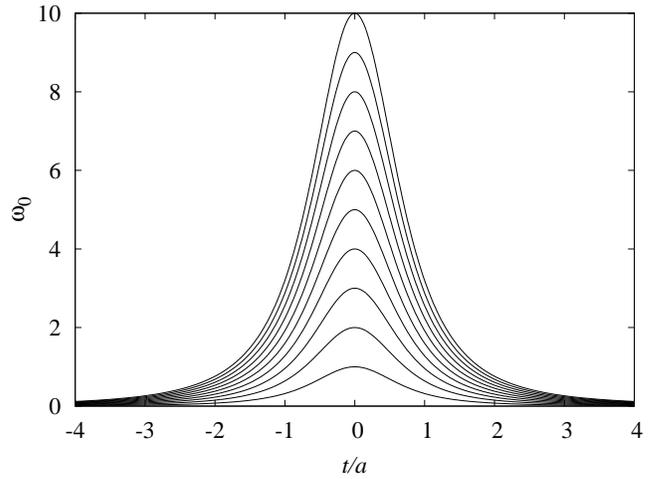}
\caption{\label{Figure2} Angular velocity of the central inertial frame $\omega_0(l,m;t)$ for $l=10$ and $m=1,2,... 10$ (from bottom to top).
The vertical axis is scaled in units of $\omega_0(10,1;0)$.}
\end{figure}
\begin{figure}[t]
\includegraphics[angle = 270, width=8.2cm, clip=true, bb=65 104 546 744]{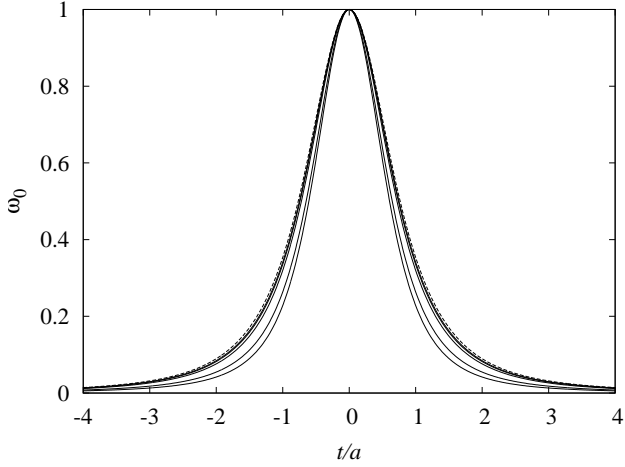}
\caption{\label{Figure3} The dependence of normalized angular velocity of the central inertial frame $\omega_0(l,1;t)/\omega_0(l,1;t=0)$ 
on the parameter $l=2,$ $3, 10, 20, 30$ (from inside to out). 
Also the function $(1+\tilde t^2)^{-3/2}$ is shown as a dashed line to indicate the limit for large $l$ (cf. Eq. (\ref{s7eq14})). }
\end{figure}

Substituting for $\kappa$ and $\lambda$ the expressions (\ref{s7eq8}) and (\ref{s2eq15}) and integrating over the radial
coordinate we obtain the angular velocity $\omega_0$ in the form
\begin{align}
\label{s7eq10}
 & \omega_0(t) = 
\frac{\tilde B_l^2}{2\pi}\,\frac {m\left (l+1\right )\left (l+2\right )}{a\left (l+3\right )l}
\times
\\\nonumber
&~~~\left[\left( U_l -V_l {\tilde t\,}^2 \right)  \left (1+{\tilde t\,}^{2}\right ) I_{l+3}^2(\tilde t) + \left (U_l+V_l {\tilde t\,}^{2}\right ) I_{l+3}^1(\tilde t)\right],
\end{align}
where
\begin{align}
\label{s7eq11}
U_l &= \left (2\,{l}^{5}+7\,{l}^{4}+4\,{l}^{3}-7\,{l}^{2}+24\,l+36\right ),
\\\nonumber
V_l &= 3\left ({l}^{4}+2\,{l}^{3}+3\,{l}^{2}+8\,l+12\right )~.
\end{align}
Here the radial integrals $I_L^M$ are defined and analyzed in Appendix C. 
Although the last result has indeed an analytic, explicit character, it is not very practical for
calculations because the integrals $I^{1,2}_{l+3}$ are determined by high-order derivatives for large $l$.
Hence, we describe their properties in detail in Appendix C.

Turning to the numerical evaluation of the angular velocity of dragging we obtain nice
``bell-shaped'' figures for $\omega_0$ for various values of $m$ at a given $l$. In Figure \ref{Figure2},
$\omega_0$ is plotted for $l=10$ and $m = 1,2,3, ... 10$.

Also, the integrals can be explicitly calculated at the moment of time-symmetry when
the dragging near the origin is maximal since the rotating waves are at nearest distance.
The integrals have simple forms also at $t = \pm a$, but these values are of no special
significance though they can serve for checking the numerical results. The central and maximal value reads
\begin{align}
\omega_0(t=0) = 
\frac{\tilde B_l^2}{4\pi}\frac{m}{a}  {\frac {\left (l+1\right )!\,\left (l+2\right )!}{l \left (2\,l+3\right)!}} U_l~.
\end{align}

Finally, at late times the analytic behavior of $I^M_L$ implies
\begin{align}
\frac{\omega_0(t)}{\omega_0(t=0)} \sim  \left| \frac{a}{t} \right|^3  + O\left(\left| \frac{a}{t} \right|^5 \right)~,
\end{align}
and in Figure \ref{Figure3} we illustrate that $\omega(t)$ for large $l$ and given $m$ can be approximated by function 
\begin{equation}
\label{s7eq14}
\omega_{l\rightarrow \infty} = \frac{ \omega_0(t=0) } {\left[1+\tilde t^2\right]^{\frac{3}{2}} } ~. 
\end{equation}

\section{Observing stars through gravitational waves}
Since our waves are sufficiently weak near infinity, we can consider static stars there and ask how they would move across the sky as seen by an observer fixed in the flat region at the origin. We shall evaluate the first-order effects of the waves on the propagation of photons, the effects due to dragging of observer's frame are of second order and will be discussed subsequently. The influence of gravitational waves on the passing photons can be computed as a perturbation of the ingoing radial null geodesic
\begin{align}
 x^{(0)\mu} = [T + \lambda, -\lambda, \theta, \varphi], ~~\lambda \in (-\infty, 0]~,
\end{align}
which describes a ray from the star with celestial coordinates $\theta, \varphi$ which reaches the center at time $T$.
The geodesic equation can be written in the form
\begin{align}
 \frac{d}{d\lambda} x^\mu =& p^{(0)\mu} +  p^{(1)\mu} + ...~,
\\
 \frac{d}{d\lambda} p^{(1)\mu} =&
 - \Gamma^{(1)\mu}_{\rho\sigma}\left(x^\nu= x^{(0)\nu}(\lambda) \right) p^{(0)\rho}p^{(0)\sigma}
\\\nonumber
 &- 2\Gamma^{(0)\mu}_{\rho\sigma}\left(x^\nu= x^{(0)\nu}(\lambda) \right) p^{(1)\rho}p^{(0)\sigma}~.
\end{align}
To solve these differential equations we use covariant components of the perturbed photon momenta for which the effects of curvilinear coordinates cancel out; we omit superscript ${}^{(0)}$ for simplicity. Then we can write
\begin{align}
&\frac{d}{d\lambda} p^{(1)}_{\theta}
=
-r^2 \frac{d}{d\lambda} p^{(1) \theta} 
- 2r p^{(1) \theta} p^r
\\\nonumber
&~~~~=
-({p^t})^{2}{h_{t\theta,t}}
-({p^r})^{2}{h_{r\theta,r}}
- {p^t}\,{p^r}\left[ {h_{r\theta,t}}+{h_{t\theta,r}} \right]
\\
&\frac{d}{d\lambda} p^{(1)}_{\varphi}
=
-r^2 \sin^2 \theta \frac{d}{d\lambda} p^{(1) \varphi}  
-2 r p^{(1) \varphi} \sin^2 \theta p^r
\\\nonumber
&~~~~=
-({p^t})^{2}{h_{t\varphi,t}}
-({p^r})^{2}{h_{r\varphi,r}}
-{p^t}{p^r}\left[ {h_{r\varphi,t}}+ {h_{t\varphi,r}} \right].
\end{align}
\begin{figure}[hb]
\begin{center}
\includegraphics[angle = 270, width=4.2cm,clip=true,bb=64 172 541 650]{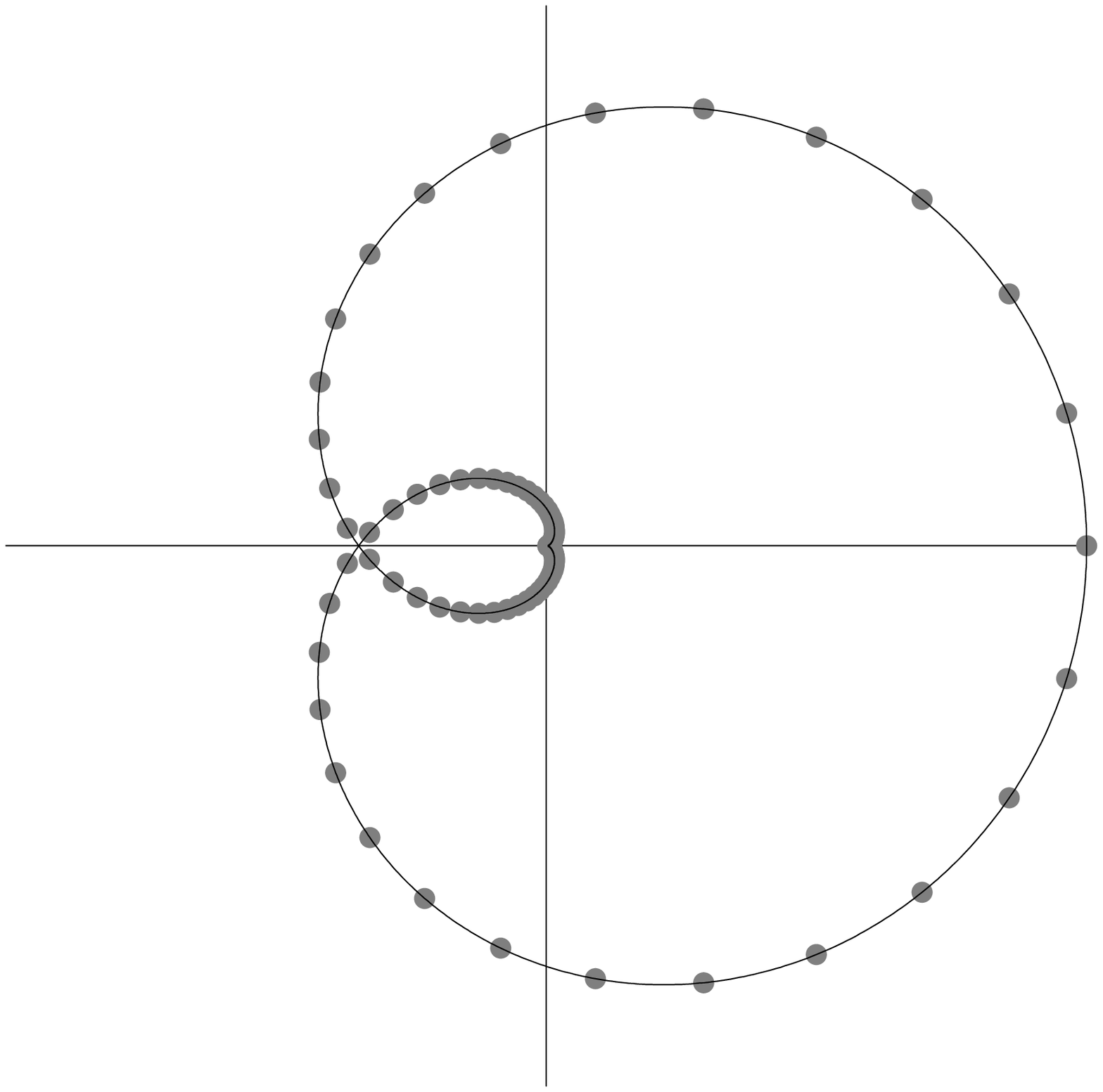}
\includegraphics[angle = 270, width=4.2cm,clip=true,bb=64 172 541 650]{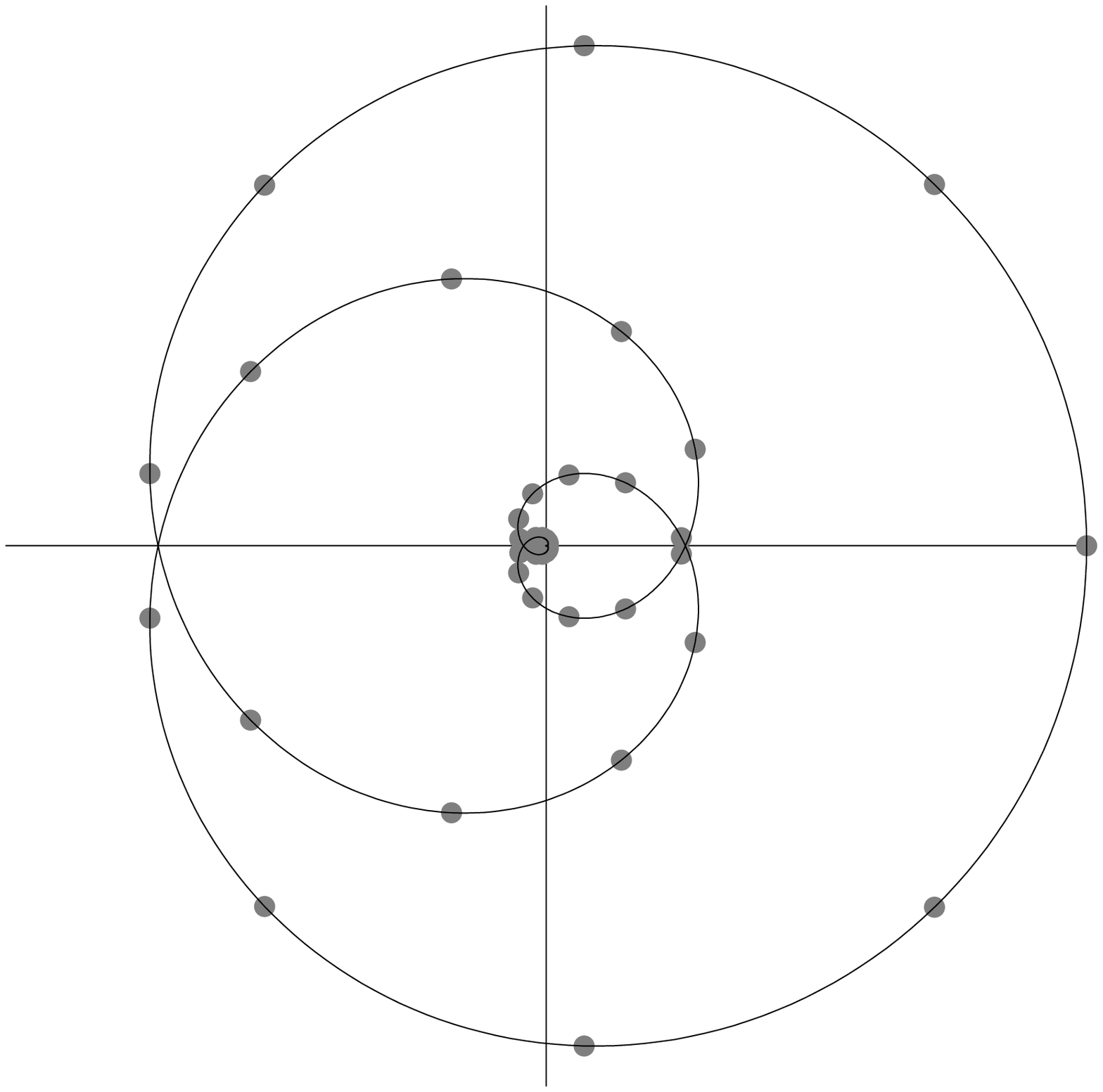}
\\
~$l=3$~~~~~~~~~~~~~~~~~~~~~~~~~~~~~~~~\;
$l=13$
\\
\includegraphics[angle = 270, width=4.2cm,clip=true,bb=64 172 541 650]{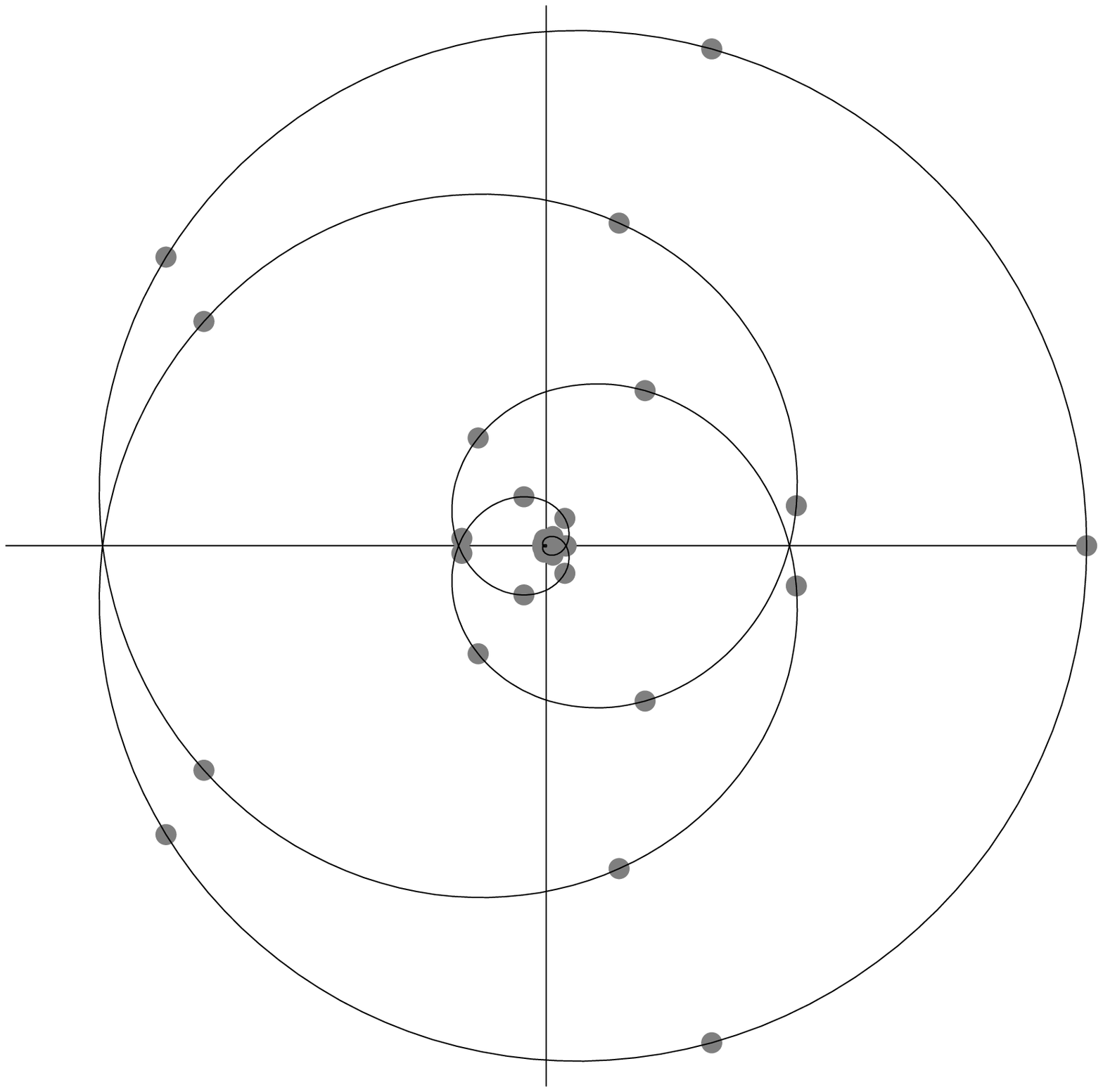}
\includegraphics[angle = 270, width=4.2cm,clip=true,bb=64 172 541 650]{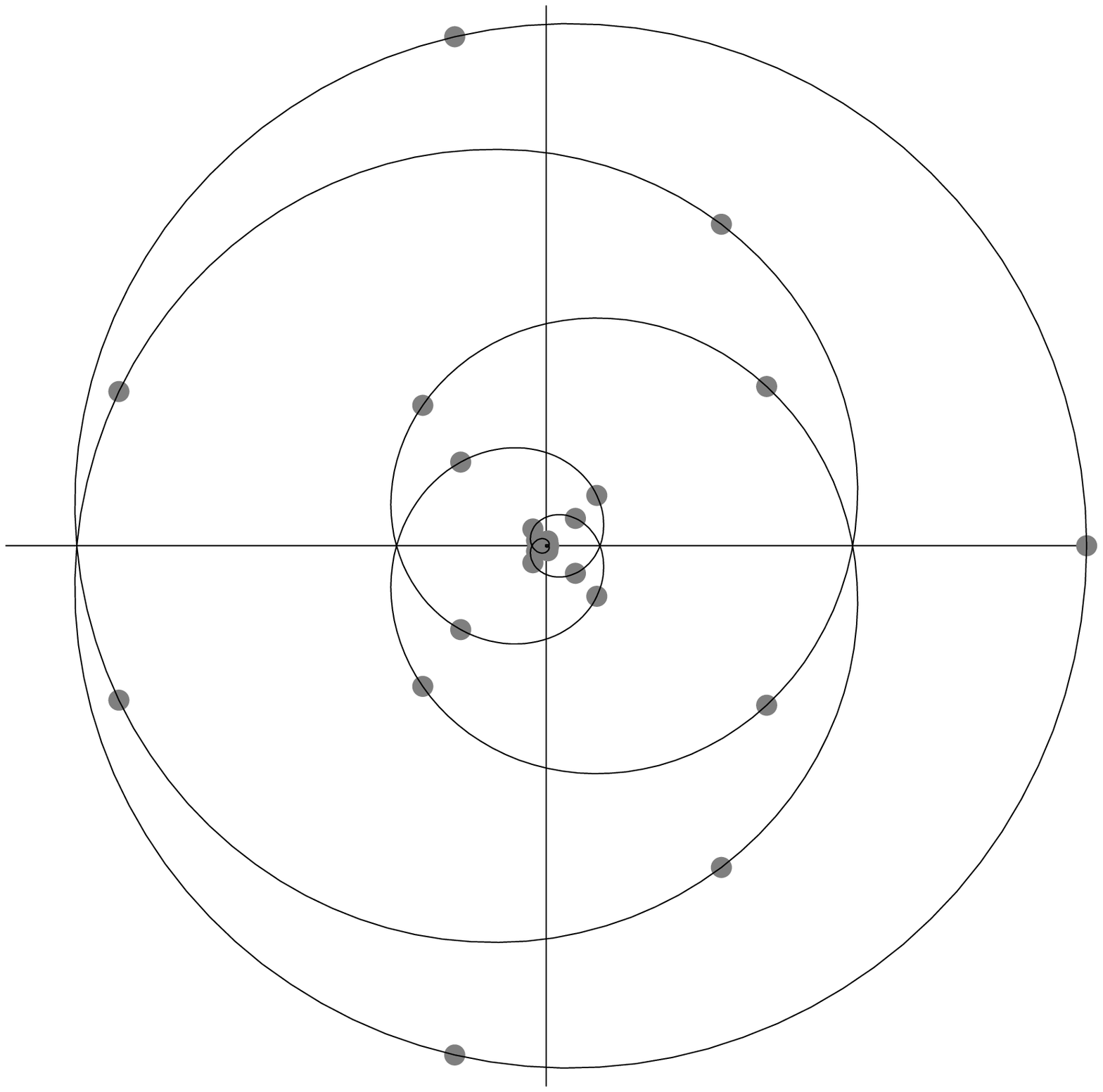}
\\
~$l=23$~~~~~~~~~~~~~~~~~~~~~~~~~~~~~~~
$l=33$
\caption{\label{Figure4}
Since light from distant stars is influenced by the gravitational waves the observed positions of the stars change.
An observer at the origin can record the apparent position of the stars on the celestial sphere on a photographic plate.
When appropriately scaled and rotated, the trajectories of all stars are the same. 
A star starts at the origin of the plate ($x=y=0$ in the planes above) and moves along closed trajectories the structure 
of which becomes more complicated with increasing $l$. At the moment of time symmetry, $\tilde t=0$, the image is located 
at maximal value of $x$ (with $y=0$).
This curve is an image of a straight line in the complex plane $z=1+is$, $s\in(-\infty,\infty)$ mapped by the function  $f(z) = z^{-l-2}$
(cf. Eq. (\ref{s8eq15})). Together with the trajectory, positions at time $\tilde t = 0,\pm 0.05,\pm 0.1,...$ are shown as circles.
}
\end{center}
\end{figure}

We see that along the radial rays (for which $\sin \theta^{(0)}=\rm const$) the covariant components of photons momenta $p^{(1)}_\theta$ and $p^{(1)}_\varphi$
can be written as integrals along the {\it un}perturbed ray and we do not need to solve the geodesic equation as set of differential equations.
Similarly, the differential equations for ${\theta}^{(1)}$ and ${\varphi}^{(1)}$ can be written as follows
\begin{align}
\frac{d}{d\lambda} {\theta}^{(1)}
&= -\frac{1}{r^2}   p^{(1)}_{\theta}
\\\nonumber
&= \frac{1}{p^{r}} \frac{d}{d\lambda} \left(\frac{1}{r}  p^{(1)}_{\theta} \right) - \frac{1}{p^{r}} \frac{1}{r} \frac{d}{d\lambda} p^{(1)}_{\theta}~,
\\
\frac{d}{d\lambda} {\varphi}^{(1)}
&= -\frac{1}{r^2\sin^2 \theta} p^{(1)}_{\varphi}
\\\nonumber
&= \frac{1}{p^{r}} \frac{d}{d\lambda} \left(\frac{1}{r\sin^2 \theta}  p^{(1)}_{\varphi} \right) - \frac{1}{p^{r}} \frac{1}{r\sin^2 \theta} \frac{d}{d\lambda} p^{(1)}_{\varphi}
~.
\end{align}
For photons which reach the origin the boundary terms arising after integration by parts vanish, so we can write
\begin{align}
\label{s8eq8}
\delta\theta(T) &= \left[ {\theta}^{(1)}\right]^{r=0}_{r=\infty}
= -\int_{-\infty}^0  \frac{1}{p^{r}} \frac{1}{r} \frac{d}{d\lambda} p^{(1)}_{\theta} d\lambda
\nonumber\\
=& -\int_{\infty}^0  \frac{1}{(p^{r})^2} \frac{1}{r} \frac{d}{d\lambda} p^{(1)}_{\theta} dr
\\\nonumber
=& \int_0^\infty \left[{h_{r\theta,t}}+{h_{t\theta,r} - {h_{t\theta,t}} -{h_{r\theta,r}} } \right]_{| t=T-r} \frac{dr}{r}
~,
\end{align}
\begin{align}
\label{s8eq9}
\nonumber
\delta\varphi(T) &= \left[ {\varphi}^{(1)}\right]^{r=0}_{r=\infty}
= -\int_{-\infty}^0 \frac{1}{p^{r}} \frac{1}{r\sin^2 \theta} \frac{d}{d\lambda} p^{(1)}_{\varphi} d\lambda
\\
=& -\int_{\infty}^0 \frac{1}{(p^{r})^2} \frac{1}{r\sin^2 \theta} \frac{d}{d\lambda} p^{(1)}_{\varphi} dr
\\\nonumber
=& \int_0^\infty \!\!\!\left[  {h_{r\varphi,t}}+ {h_{t\varphi,r}}  -{h_{t\varphi,t}} -{h_{r\varphi,r}} \right]_{| t=T-r}\frac{dr}{r\sin^2 \theta}
~.
\end{align}
Since now we work in the linear approximation  to find the light deflection we can go back to the simpler, complex-valued formulas (\ref{s2eq10}) and (\ref{s4eq14}-\ref{s4eq19}).
When we introduce amplitudes 
\begin{align}
\Delta\theta &=
 {B_l N^m_l} 2(l-1)! \frac{m P^m_l(\cos \theta) }{\sin \theta},
\\
\Delta\varphi &= 
- {B_l N^m_l}2(l-1)! P'^m_{~l}(\cos \theta),
\end{align}
the integrals (\ref{s8eq8}) and (\ref{s8eq9}) simplify to 
\begin{align}
\delta\theta(T) &= {Re}\big[\Delta\theta\; \rho(a+iT) \big],
\\
\delta\varphi(T) &= {Re}\big[\Delta\varphi\; i \rho(a+iT) \big],
\end{align}
where the complex function $\rho(a+iT)$ reads
\begin{align}
\rho&= -i \;  e^{im\varphi} \frac{l(l+1)(l+2)}{\left (a+ i T\right )^{l-1}}\int^{\infty}_0{\frac {a^{l+2}\;{(2r)}^{l-1}}{\left (a+ i T-2\, i r\right )^{l+3}}}~dr
\nonumber\\
&= -  e^{im\varphi} i^{l+1} \frac{1}{\left(1+i\frac{T}{a}\right)^{l+2}}~.
\end{align}

If we assume a telescope pointed towards the star's initial position at $T\rightarrow -\infty$,
the quantities $\delta \theta$ and $\sin \theta \delta \varphi$ are proportional to star's coordinates 
on telescope's photographic plate; star's trajectory can be described by a simple formula
\begin{align}
  \label{s8eq15}
  \frac{\delta\varphi}{\Delta\varphi} +  i \frac{\delta\theta}{\Delta\theta} = \frac{i^l\;e^{im\varphi}}{\left(1+i \frac{T}{a} \right)^{l+2}}~.
\end{align}
We see that the same image is replicated for all stars (Figure \ref{Figure4}).
Depending on the star's position the image is rotated by $e^{ i m\varphi}$ and
then scaled by factors $\Delta\theta$ in the latitudinal and by $\Delta\varphi$ in the longitudinal direction. Surprisingly, the images are time-symmetric with the maximal deflection occurring at $T=0$ despite the fact that the deflection is calculated using retarded integrals (\ref{s8eq8}) and (\ref{s8eq9}). 

Since the calculation of the second-order metric everywhere would be very complicated we cannot, of course, describe all second-order effects. 
However we can describe the second-order effect of the total rotation of the central observer's telescope due to the dragging of his inertial frames given above, cf. (\ref{s7eq10}), (\ref{s7eq11}). Since the pulse starts and ends at infinity at $t=\pm \infty$, the total rotation of the telescope is clearly gauge invariant.
As a consequence, the curves in Figure \ref{Figure4} will no longer be closed. 
The stellar initial position at $T \rightarrow -\infty$ will differ from the final position at $T \rightarrow +\infty$ 
by an amount proportional to the area under the graph of $\omega_0(t)$.

\section{Concluding remarks}
The results obtained here extend significantly our past studies \cite{BKL08cqg1,BKL08cqg2} of gravitational waves with the translational symmetry that rotate about the axis of cylindrical symmetry and cause a rotation of local inertial frames at the axis. In the present work we investigated the effects of the rotating gravitational waves in a general, asymptotically flat situation. The main conclusions following from these studies demonstrate explicitly that any statement of Mach's principle that attributes
all dragging of inertial frames solely to the distribution of energy and momentum of matter, characterized by $T_{\mu\nu}$, as the origin of inertia is false --
gravitomagnetic effects are also caused by gravitational waves satisfying Einstein's field equations in vacuum for which a local energy and momentum cannot in general be defined.

We found that the dragging of inertial frames at the origin occurs {\it instantaneously}, with no time-delay  depending on the position of the pulse. The dipole perturbations  determining the dragging follow from the constraint equation  which is elliptic (see equation  (\ref{s6eq11})). Gravity is a tensor field, so gravitational waves  cannot have a dipolar part. This is the case with our first-order rotating pulse. It is the non-linearity of Einstein's field equations  causing a self-interaction  of the waves which produces an effective source for the second-order perturbations. For a general multipole moment $l\ge 2$, the resulting equations  are again wave-type equations  indicating that the effects of the first-order terms cause the second-order terms in a non-instantaneous way. However, inertial frames  at the origin will only be influenced by the dipole perturbations  since waves decay as $r^l$ at the origin. And for the {\it dipole} $(l=1)$ second-order perturbations, the equation  becomes elliptical,
 with the source term given by a component of the second-order Ricci tensor $R_{t \varphi}^{(2)}$ quadratic in the first-order metric and its derivatives.

The instantaneous character of the dragging of inertial frames  by angular momentum  of sources implied by the Einstein constraint equation  can, in fact, be seen in diverse situations, including those in cosmological perturbation  theory. We encountered the effect in the study of local inertial frames  affected by the rotations beyond the cosmological horizon \cite{BLK1}, \cite{BLK2}, or in the general analysis of ``rotations and accelerations" of local inertial frames  in the linearly perturbed Friedmann-Robertson-Walker  spacetimes  \cite{BKL07}. In the cosmological context we introduced desirable gauges - we call them the ``Machian gauges" - as those in which local inertial frames  can be determined instantaneously from the distributions of energy  and momentum in the universe by means of the perturbed Einstein constraint field equations. The ``uniform-Hubble-expansion-gauge" or the ``minimal-shear hypersurface" gauge combined with ``the minimal distortion" gauge on the spatial metric are examples of the 
Machian gauges; the synchronous gauge or the generalized Lorentz-de Donder (harmonic) gauge are not. (See \cite{BKL07} for detailed analysis and preferences of the Machian gauges.)

Likewise, in the context of gravitational waves  we should bear in mind that such physical effects like the dragging of inertial frames are of a global nature and require the introduction of suitable coordinates in which the waves are described\footnote{This point has been emphasized also in contributions and discussions in the excellent book \cite{BP} (see, for example, remarks by D. Brill, p. 213).}.

In the present work we also investigated for the first time, as far as we are aware, the effects of the rotating gravitational waves on the propagation of light from distant fixed stars at (asymptotically flat) infinity. 

There are various open problems in these themes which deserve further study. In order to have more natural understanding of the effect of rotating gravitational waves one should consider the waves within a cosmological (not asymptotically flat) context. Although in \cite{BKL07} we emphasized the role of the instantaneous Machian gauges, it would be also interesting to study these effects, in particular those connected with gravitational waves, in other type of gauges like the 
Lorenz-de Donder gauge (cf. Appendix D in \cite{BKL07}). 

Of course, any simple model problem illustrating the Machian effects {\it exactly} within general relativity, like that employing the conformal stationary metrics in our recent note on the linear gravitational dragging \cite{cqg12}, deepens considerably our understanding of such effects.


\section*{Acknowledgments}
J.B. and J.K acknowledge the hospitality of the Institute of Astronomy in Cambridge. J.B. and T.L. acknowledge the partial support from the Grant GA CR 202/09/00772 of the Czech Republic and the Grant No MSM0021620860 of the Ministry of Education.

\appendix
\section{Integrals over products of Legendre functions}
The following integrals over $\theta$ or $\mu=\cos\theta$ are needed to evaluate functions $h_0(t,r)$ 
in (\ref{s6eq13}) in terms of radial integrals over $g(t,r)$ given by (\ref{s6eq12}) and (\ref{appBeq3}). 
Most of them can be found in \cite{GR}, two follow from simple integration by part. These are ($0<m\le l$):
\begin{align}
\label{A1}
\int_{-1}^1(P^m_l)^2d\mu&=\frac{2}{2l+1}\frac{(l+m)!}{(l-m)!},
\\
\int_{-1}^1\frac{(P^m_l)^2}{1-\mu^2}d\mu&=\frac{1}{m}\frac{(l+m)!}{(l-m)!},
\\
\label{A2}
\int_{-1}^1\mu P^m_l{P^m_l}'d\mu&= - \frac{1}{2l+1}\frac{(l+m)!}{(l-m)!},
\end{align}
\begin{align}
\int_{-1}^1\!\!\! (1-\mu^2)({P^m_l}')^2d\mu&= \frac{(l+m)!}{(l-m)!}\left[  \frac{2l(l+1)}{2l+1} - m  \right]\!.
\end{align}
\section{Calculating and averaging the second-order Ricci tensor component}
To compute $R^{(2)}_{t\varphi}[h^{(1)},h^{(1)}]$ we use metric perturbations (\ref{s7eq4}), (\ref{s7eq5}) which in coordinates $x^\mu = \{t,r,\theta,\varphi\}$ 
yield a linearized metric with odd-parity perturbations on a flat background. We start from the general form
\begin{align}
g_{\mu\nu} = \left[
  \begin{array}{cccc}
    1& 0& h_{t \theta}& h_{t \varphi}\\
     0& -1& h_{r \theta}& h_{r \varphi}\\
     h_{t \theta}& h_{r \theta}& -r^2& 0\\
     h_{t \varphi}& h_{r \theta}& 0& -r^2 \sin^2 \theta
  \end{array} \right]~.
\end{align}
With the metric functions given by  (\ref{s7eq4}), (\ref{s7eq5}) the components of the Ricci tensor in the first-order approximation yield zero either identically or due to the fact that $\psi_{lm}$ solves the wave equation (\ref{s2eq1}).
The only second-order component we need is 
\begin{align}
\label{appBeq2}
\nonumber
& - R^{(2)}_{t \varphi}  = {\frac {1}{2{r}^{2}}}\Big(
  h_{t \theta,r} h_{r \varphi,\theta} 
 -h_{t \varphi,\theta} h_{r \theta,r}
 -h_{t \varphi,r} h_{r \theta,\theta}
\\\nonumber
& +h_{t \theta} ( h_{t \varphi,t \theta} + h_{t \theta,\varphi t})
 + h_{r \varphi,t} h_{r \theta,\theta}
 -h_{r \theta,\varphi} h_{r \theta,t}  
\\ \nonumber
&+h_{r \theta} ( h_{t \theta,\varphi r} +  h_{r \varphi,t \theta}  - 2 h_{t \varphi,r \theta}  -2h_{r \theta,\varphi t} )
+h_{t \theta,\varphi} h_{r \theta,r}
\Big)
\end{align}
\begin{align}
&
+{\frac{1}{ r^2}\frac {\cos \theta}{\sin \theta}}
\left (
    h_{t \varphi} h_{t \theta,t}
 -  h_{r \varphi} h_{t \theta,r}
 -  h_{t \theta} h_{t \varphi,t}
\right )
\\ \nonumber
& + {\frac {1}{2 {r}^{2} \sin^2\theta}}
\left( 
 2 h_{t \varphi} h_{t \varphi,\varphi t}
 -h_{r \varphi} ( h_{t \varphi,\varphi r} + h_{r \varphi,\varphi t})
\right)
\\ \nonumber
&+{\frac {1}{{r}^{3}}
h_{r \theta} \left( h_{t \varphi,\theta} - h_{t \theta,\varphi} \right) }
+
{\frac{1}{2 r^2}\frac {\cos \theta}{\sin \theta}}
h_{r \theta}
\left ( h_{r \varphi,t} + h_{t \varphi,r} \right )
~.
\end{align}
This simplifies to (\ref{s7eq6}) when perturbations in the form (\ref{s7eq4}), (\ref{s7eq5}) are introduced. When integration over the spherical angles $\varphi$ and $\theta$
is performed, we get (\ref{s7eq9}). At this moment we use functions $\kappa$  and $\lambda$ given in (\ref{s7eq8}) and (\ref{s2eq15}) and obtain 
\begin{align}
\label{appBeq3}
& \int_0^\infty \!\!\!\int R_{t \varphi} d\Omega \frac{dr}{r}=
\int_0^\infty
\frac {2\tilde B_l ^2\,m\left (l+1\right ){a}^{2\,l+5}{r}^{2\,l-1}~dr}
       {\left [r_-^{4}+4\,{a}^{2}{t}^{2}\right ]^{l+3}}
\times
\nonumber\\&~~~~~
\Bigg\{r_+^2\left (r_-^4+4\,{a}^{2}{t}^{2}\right )
\left ({l}^{4}+2\,{l}^{3}+3\,{l}^{2}+12\,l
\right )
\\\nonumber
&~~~~~~~~
+4\,\left ({t}^{2}+{a}^{2}\right )
\left [
3r_+^{4}-4\,{r}^{2}\left (2\,{a}^{2}+2\,{r}^{2}+{t}^{2}\right )
\right ]
\\\nonumber&~~~~~~~~
+8\,{r}^{2}l \left[
a^4 + r^4 + 2 a^2 t^2-t^4
-2\left (l+4\right )a^2 r_+^2
\right]
\Bigg\},
\end{align}
where $r_\pm^2={r}^{2}+{a}^{2}\pm{t}^{2}$.
After the numerator is expanded in powers of $r$, using $\tau=t/a$, this integral can be converted into 
a linear combination of integrals $I^M_L$ defined in Appendix C: 
\begin{align}
\label{appBeq4}
\nonumber
&  \int_0^\infty \!\!\! \int R_{t \varphi}~ d\Omega~ \frac{dr}{r}=
\frac{
\tilde B_l^2
m\left (l+1\right )}{a}
\times
\nonumber\\
&
\Bigg[
 { {Z_l I^{0}_{l+3}}}
+ { {\left (Z_l+8\,l+12\right )\left ({\tau}^{2}+1\right )^{3} I^3_{l+3}}}
\\\nonumber
&+{{\left ({\tau}^{2}+1\right )\left [3Z_l-16(l+1)^2+8-\left(Z_l-8\right ){\tau}^{2}\right ] I^2_{l+3}}}
\\\nonumber
&+{{\left [
3Z_l-(4l+5)^2+5
-\left(Z_l+8\,l+20\right){\tau}^{2}
\right ] I^1_{l+3} }}
\Bigg]~,
\end{align}
where $Z_l=l\left ({l}^{3}+2\,{l}^{2}+3\,l+4\right )$.
Using identities (\ref{appCeq15}) in Appendix C the number of integrals $I^M_L$ can be reduced down to two, as is shown in the final form of the formula for $\omega_0$ in (\ref{s7eq10}).

\section{Radial integrals $I^M_L, J^M_L$}
To evaluate the metric function $h_0$ at a general radius $r$ we need indefinite integrals in (\ref{s6eq13}); 
to determine the rotation of the inertial frame  at the origin we have to evaluate the definite integral in (\ref{appBeq3}). 
If we introduce dimensionless variables 
$\zeta = {r^2}/{a^2}$, and $\tau = {t}/{a}$,
the definite integrals  we need to evaluate simplify into
\begin{equation}
I^M_L(\tau)= 
\int_0^\infty\frac{\zeta^{L-M-1}}{\left[(1+\tau^2)^2+2(1-\tau^2)\zeta+\zeta^2\right]^L}d\zeta,
\label{appCeq1}
\end{equation}
where $L$ will be $l-3$, $M$ is a small integer. We first put
\begin{equation}
x=\frac{\zeta}{1+\tau^2}~~~,~~~\beta=\frac{1-\tau^2}{1+\tau^2},
\label{Cc2}
\end{equation}
so that
\begin{equation}
I^M_L=\frac{(1+\tau^2)^{L-M}}{(1+\tau^2)^{2L}}\int_0^\infty \frac{x^{L-M-1}}{(1+2\beta x+x^2)^{L}}dx.
\label{appCeq3}
\end{equation}
To evaluate this integral we start from the following simple integral (with $\alpha$=const parameter)
\begin{align}
\label{Cc4}
H_0&=\int_0^\infty\!\!\!\! \frac{dx}{\alpha+2\beta x+x^2}
\\\nonumber
&= \frac{1}{\sqrt{\alpha-\beta^2}}\textstyle{\left(    \frac{\pi}{2} - \arctan\frac{\beta}{\sqrt{\alpha-\beta^2}}     \right)}.
\end{align}
Setting then
\begin{equation}
H_M\equiv\int_0^\infty \!\!\frac{dx}{(\alpha+2\beta x+x^2)^{M+1}}={\left(-\frac{\partial}{\partial\alpha}  \right)^{\!M}\frac{H_0}{M!}},
\label{appCeq5}
\end{equation}
we can find the integral which appears in (\ref{appCeq3}) by evaluating
$(L-M-1)$-th partial derivative of $H_M(\alpha,\beta)$
 with respect to $\beta$,
so the original integral turns out to be
\begin{eqnarray}
I^M_L
 =
\frac{2^M (1+\tau^2)^{-L-M}}{(-2)^{L-1}(L-1)!}
\left. \frac{\partial^{L-1}H_0}{\partial^M\alpha\; \partial^{L-M-1}\beta} \right|_{\alpha=1}.
\end{eqnarray}
Since it is inconvenient to use this general form we will write the two integrals which appear in (\ref{s7eq10}) explicitly:
\begin{align}
\label{appCeq7}
 I^1_{l+3} &= \frac{(1+\tau^2)^{-l-4}}{2^{3l+6} (l+2)!}   \left( \frac{(\tau^2+1)^2}{\tau} \frac{d}{d\tau} \right) ^{l+1} \Pi_{1}~,
\\\nonumber
\Pi_{1}&=\frac{(\tau^2+1)^3}{\tau^3} \arctan \tau + \frac{\tau^4-1}{\tau^2}~, 
\\
\label{appCeq8}
I^2_{l+3} &= \frac{(1+\tau^2)^{-l-5}}{2^{3l+6} (l+2)!}   \left( \frac{(\tau^2+1)^2}{\tau} \frac{d}{d\tau} \right) ^{l} \Pi_2~,
\\\nonumber
\Pi_{2}&=3\frac{(\tau^2+1)^5}{\tau^5} \arctan \tau +\frac{\tau^4-1}{\tau^4} (3\tau^4+14\tau^2+3)~.
\end{align}

The expressions for these integrals become much simpler at $\tau=0$ (and at $\tau=\pm 1$ which, however, we will not discuss further). 
At the moment of time symmetry the denominator of the integral (\ref{appCeq1}) simplifies to $(1+\zeta)^2$, and we get integrals of the form
\begin{equation}
J^M_L=\int_0^\infty \frac{x^M}{(1+x)^L}dx.
\end{equation}
Assuming $L-M\ge2$ (which is satisfied in the cases we need), we can easily integrate by parts and get
\begin{equation}
 J^M_L=\left[ -\frac{1}{ L-1} \frac{x^M}{(1+x)^{L-1}}   \right]^\infty_0+\frac{M}{L-1}J^{M-1}_{L-1},
\end{equation}
where the first term vanishes if $L-M\ge 2$. Continuing with integration by parts we arrive at the simple result:
\begin{equation}
J^M_L=\frac{M!(L-M-2)!}{(L-1)!}.
\end{equation}
This yields 
\begin{equation}
I^M_L(\tau=0) = \frac{(L-M-1)!\,(L+M-1)!}{(2L-1)!}.
\end{equation}
For large $|\tau|$ (\ref{appCeq7}) and (\ref{appCeq8}) provide
\begin{align}
I^1_{l+3}(\tau) 
&\simeq 
\frac{(2l+3)!!}{2^{3l+9}(l+2)!}\frac{\pi}{|\tau|^{3}} + O\left(\frac{1}{|\tau|^5}\right),
\end{align}
\begin{align}
I^2_{l+3}(\tau)
&\simeq 
\frac{(2l+3)!!}{2^{3l+9}(l+2)!}\frac{\pi}{|\tau|^{5}} + O\left(\frac{1}{|\tau|^7}\right).
\end{align}

Even though (\ref{appBeq4}) contains integrals $I^M_{l+3}$ with $M=0,1,2,3$, only two of them are ``independent'' because of
the fact that $\int_0^\infty f'(x) dx =0$ for functions satisfying $f(0)=f(\infty)$ implies 
\begin{align}
\label{appCeq15}
&\left(L-M-1\right)\left(1+\tau^2\right)^2 I^{M+1}_L 
\\\nonumber
&~~~~~ - 2 M \left(1-\tau^2\right)\,I^{M}_L
-\left(L+M-1\right)\,I^{M-1}_L = 0~.
\end{align}


\end{document}